\pdfoutput=1%
\documentclass[a4paper,aip,american,citeautoscript,floatfix,jcp,longbibliography,pdftex,superscriptaddress,%
reprint,twocolumn]{revtex4-1}
\usepackage{amsfonts,amsmath,amssymb}%
\usepackage{graphicx}%
\usepackage[utf8]{inputenc}%
\usepackage[T1]{fontenc}
\usepackage{textcomp}%
\usepackage[textsize=footnotesize]{todonotes}%
\usepackage{xspace}
\usepackage[breaklinks=true]{hyperref}%
\usepackage{breakurl}%
\usepackage{hypernat}%
\usepackage{soul}%
\usepackage{mathtools}

\newcommand{\partOne}{(I)\xspace}

\newlength{\figwidth}%
\newlength{\figwidthlarge}%
\newlength{\figwidthsmall}%
\newcommand{\mathematicaFig}{0.38}

\usepackage{rotating}
\usepackage{ mathrsfs } 
\usepackage{braket}
\usepackage{ upgreek }
\usepackage{siunitx}

\newcommand{\ri}[1]{ \mathrm{_{#1}}}
\newcommand{\kd}[2]{\delta_{{#1}{#2}}}

\newcommand{\spht}[3]{\mathrm{T}^{#1}_{#2}\left[#3\right]}

\newcommand{\WignerD}[6]{\mathcal D^{#1} _{{#2}{#3}} (#4,#5,#6)} 

\newcommand{\WignerDop}[3]{\mathcal D^{#1} _{{#2}{#3}} } 

\newcommand{\tj}[6]{ \begin{pmatrix}
  #1 & #2 & #3 \\
  #4 & #5 & #6 
 \end{pmatrix}}
 \newcommand{\sj}[6]{ \begin{Bmatrix}
  #1 & #2 & #3 \\
  #4 & #5 & #6 
 \end{Bmatrix}}
 \newcommand{\ninej}[9]{ \begin{Bmatrix}
  #1 & #2 & #3 \\
  #4 & #5 & #6 \\
  #7 & #8 & #9
 \end{Bmatrix}}
\newcommand{\ntwopXstate}{X$^2\Sigma_\mathrm{g}^+$\xspace}

\newcommand{\ntp}{N$_{2}^{+}$\xspace}
\newcommand{\ntwoAppState}{a$^{\prime \prime} \, ^{1} \Sigma_\mathrm{g}^{+}$\xspace}

\newcommand{\np}{n^+}
\newcommand{\vp}{v^+}
\newcommand{\Jp}{J^+}
\newcommand{\Fp}{F^+}
\newcommand{\MJp}{M^+_J}
\newcommand{\MJ}{M_J}

\newcommand{\MFp}{M_F^+}
\newcommand{\MNp}{M^+_N}
\newcommand{\MSp}{M^+_S}
\newcommand{\Np}{N^+}
\newcommand{\Sp}{S^+}
\newcommand{\Lp}{\Lambda^+}

\newcommand{\Ip}{I^+}

\newcommand{\MIp}{M_I^+}
\newcommand{\CG}[6]{C_{#1 #2 #3 #4} ^{#5 #6}}
\newcommand{\bBetaJ}{(b$_{\beta_J}$)\xspace}

\newcommand{\upr}{w}

\newcommand{\formsphtMolFixed}[2]{\mathrm{T'}^{#1}_{#2}}
\newcommand{\Me}{M\ri e}
\newcommand{\Mg}{M\ri g}

\newcommand{\oig}{ \omega_{\mathrm{ig}} }

\newcommand{\oievnsrg}{     \omega _{  \mathrm{i}_\mathrm{ev}  \mathrm{i}_\mathrm{nsr}   \mathrm{g}   }     }

\newcommand{\esigma}{\mathbf{e}_\sigma}
\newcommand{\exPolSF}{\sigma}	
\newcommand{\exPolMFone}{\tau_{1}}	
\newcommand{\exPolMFtwo}{\tau_{2}}	
\newcommand{\exPolMFoneNoIndex}{\tau}	
\newcommand{\kEx}{\kappa}	
\newcommand{\SppTop}{S_{'' \leftrightarrow '}}
\newcommand{\polarisationAngle}{\alpha}
\newcommand{\coloneqq}{\mathrel{\vcentcolon\mkern-1.2mu=}} 

\newcommand{\myhat}[1]{#1}

\begin{document}
\title{Fine- and hyperfine-structure effects in molecular photoionization: II. Resonance-enhanced multiphoton ionization and hyperfine-selective generation of molecular cations}%
\author{Matthias Germann}
\author{Stefan Willitsch}\email{stefan.willitsch@unibas.ch}
\affiliation{Department of Chemistry, University of Basel, Klingelbergstrasse 80, 4056 Basel, Switzerland}
\date{\today}%
\begin{abstract}
Resonance-enhanced multiphoton ionization (REMPI) is a widely used technique for studying molecular photoionization and producing molecular cations for spectroscopy and dynamics studies.
Here, we present a model for describing hyperfine-structure effects in the REMPI process and for predicting hyperfine populations in molecular ions produced by this method. This model is a generalization of our model for fine- and hyperfine-structure effects in one-photon ionization of molecules presented in the preceding companion article\cite{germann16cPROVISORISCH_arXiv}. This generalization is achieved by covering two main aspects: (1) treatment of the neutral bound-bound transition including hyperfine structure that makes up the first step of the REMPI process and (2) modification of our ionization model to account for anisotropic populations resulting from this first excitation step.
Our findings may be used for analyzing results from experiments with molecular ions produced by REMPI and may serve as a theoretical background for hyperfine-selective ionization experiments.
\end{abstract}
\maketitle%


\section{Introduction}

Resonance-enhanced multiphoton ionization (REMPI)\hspace{0cm}---\hspace{0cm}ionization of atoms or molecules by several photons via resonant excitation of the neutral precursor species\hspace{0cm}---\hspace{0cm}has become a well-established method to study the photoionization of molecules \cite{ashfold94a,ellis05a} and produce molecular cations for dynamics and spectroscopy experiments \cite{tong12a,germann14a} over the last decades.
Multiphoton ionization avoids the need for vacuum-ultraviolet radiation, while resonant ionization, i.e., ionization via an excited state of the neutral precursor molecule, generally increases the photoionization yield and improves the selectivity of the ionization process. In particular, selection and propensity rules governing the REMPI process may be exploited for rotational-vibrational state-selective production of molecular cations \cite{mackenzie95a,tong10a,tong11a}.

In the preceding article \cite{germann16cPROVISORISCH_arXiv}, cited as \partOne below, we have presented a model for fine-structure-(fs) and hyperfine-structure-(hfs)-resolved photoionization intensities in direct, one-photon ionization of molecules. Here, we extend our model to cover two-color multiphoton ionization.

We focus on the [2+1'] REMPI scheme, i.e., excitation by a two-photon transition from the neutral-ground to a neutral-excited state followed by one-photon ionization of the neutral-excited state. The method presented, however, is general and may be extended to other two-color REMPI schemes such as [1+1'].

\section{General considerations}

\begin{figure}[btp]
\begin{center}
\includegraphics{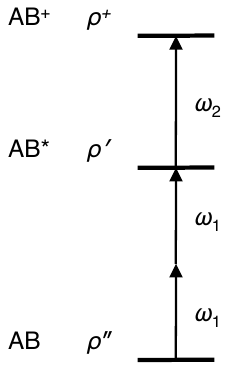}
\caption{
{\bf Schematic view of a [2+1'] resonance-enhanced multiphoton ionization process:} the neutral, electronic-ground-state molecules AB are excited by absorption of two photons at an angular frequency $\omega_1$ yielding excited molecules AB*. These are then ionized by absorption of a third photon at a different frequency $\omega_2$, resulting in the molecular ions AB$^+$. The populations associated with AB, AB* and AB$^+$ are denoted $\rho''$, $\rho'$ and $\rho^+$, respectively.
}
\label{fig:generic_REMPI}
\end{center}
\end{figure}

The scheme of a [2+1'] REMPI process is illustrated in Fig.~\ref{fig:generic_REMPI}:
A neutral diatomic molecule AB is first excited from the electronic ground state to a neutral electronically or vibrationally excited state (AB*) by absorption of two photons at the (angular) frequency $\omega_1$. Thereafter, the molecule is ionized by absorption of a third photon at a different frequency $\omega_2$ forming the molecular ion AB$^+$.
Following this picture, we describe the [2+1'] REMPI process as a sequence of two independent steps: a transition from the neutral electronic-ground-state molecule AB to the neutral excited molecule AB* and a subsequent ionization of this molecule yielding the molecular ion AB$^+$.

For the excitation step (AB $\rightarrow$ AB*), we will develop a model for hfs-resolved two-photon transitions between bound states. Using this model, we calculate the excitation rate $R(\text{AB} \rightarrow \text{AB*})$ and hence the relative rotational and hyperfine populations $\rho'$ of excited molecules AB*. The ionization of the excited molecules is then described by our ionization model developed in \partOne, with the excited state population $\rho'$ used in lieu of the thermal ground state population.

The excitation of the neutral molecules AB to AB* by polarized radiation leads to an anisotropic population $\rho'$, meaning that the several Zeeman states of a fs  or hfs level in the excited state are unequally populated.\cite{allendorf89a,reid91a} Since for our photoionization model presented in \partOne, isotropic populations have been implicitly assumed, we need to adapt that model for  anisotropic excited populations of the neutral present in REMPI.

Our manuscript is structured as follows: In Sec.~\ref{sec:Excitation_step}, we discuss the excitation step, i.e., we develop a model for hyperfine structure effects in the initial two-photon excitation transition. Besides the application in our [2+1'] REMPI model, the theory developed in that section is also generally applicable to hyperfine-structure-resolved two-photon bound-bound transitions.

In Sec.~\ref{sec:Ionization_step}, we discuss the ionization step, i.e., we develop the above-mentioned adaptations of our ionization model from (I) for ionization of an anisotropically populated neutral state.

In Sec.~\ref{sec:REMPI}, we then combine these results to a complete model for the [2+1'] REMPI process.
Moreover, we shortly indicate how to adapt our model to other REMPI processes besides the [2+1'] scheme discussed here.

The implications of our model are shown in Sec.~\ref{sec:Application} using the [2+1'] REMPI of N$_2$ via the neutral excited \ntwoAppState state as a representative example.

Finally, we summarize our findings in Sec.~\ref{sec:Conclusions}.


\section{Excitation step: Hfs-resolved non-resonant two-photon transitions}
\label{sec:Excitation_step}

Two- and multiphoton transitions in diatomic molecules have been discussed in several previous publications, e.g., by Bray and Hochstrasser \cite{bray76a}, Ma\"inos \cite{mainos86a}, Lefebvre-Brion and Field \cite{lefebvre-brion04a} as well as Hippler \cite{hippler99a}.
These treat the ``G\"oppert-Meyer mechanism'' first described in Refs.~\onlinecite{goeppert-mayer29a,goeppert-mayer31a}. Here, we will extend these treatments to hyperfine-structure-resolved transitions. 

\begin{figure*}[tbp]
\begin{center}
\includegraphics{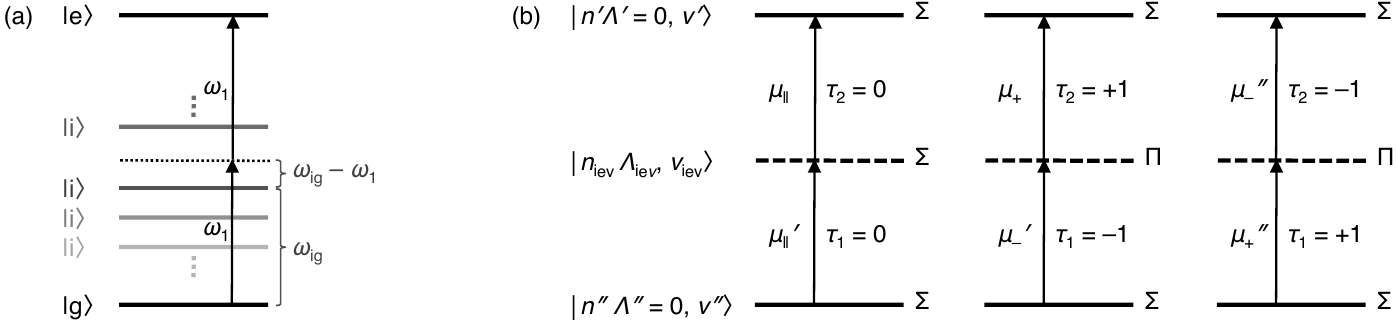}
\caption{\textbf{Mechanism of two-photon transitions: (a)} Two-photon transitions are described by virtual one-photon transition routes connecting the ground state $\Ket{\mathrm g}$ to an excited state $\Ket{\mathrm e}$ via an intermediate state $\Ket{\mathrm i}$. The two-photon line strength is given by a  weighted sum of one-photon transition routes via different intermediate states $\Ket{\mathrm i}$. Their weights---illustrated here by different grey tones---are determined by the inverse of the energy mismatch $\hbar(\oig - \omega_1)$ between the photon energy $\hbar\omega_1$ and the transition energy $\hbar\oig$. \textbf{(b)} Two-photon transitions between two $\Sigma$-states may occur via $\Sigma$ or $\Pi$ intermediate states. Transition routes are labeled by the effective electric-dipole matrix element ($\mu_{||} \mu'_{||}$, $\mu_{+}\mu'_{-}$, $\mu''_{-}\mu''_{+}$, see Eq.~\eqref{eq_two-photon_dipole_moments}) and the relevant spherical tensor component of the electric-dipole operator in the molecule-fixed frame ($\tau_{1,2} = 0, \pm 1$).
}
\label{fig:two-photon_mechanism}
\end{center}
\end{figure*}

The transition rate $R_{\mathrm{g} \rightarrow \mathrm{e}}$ for the excitation of the molecule from the ground state $\ket{\mathrm{g}}$ to the excited state $\ket{\mathrm{e}}$ is expressed as a product of the radiation intensity $I_{0}$ and the two-photon line strength $S\ri{ge}$:\cite{hippler99a}
\begin{equation}
R_{\mathrm{g} \rightarrow \mathrm{e}} \propto (I_{0})^{2} S\ri{ge} .
\end{equation}
Note that for two-photon transitions, the radiation intensity enters squared in the two-photon transition rate.

According to the G\"oppert-Meyer mechanism, the ground and excited state are connected by two off-resonant one-photon transitions via intermediate states. As excitation is possible via different intermediate states, all possible virtual transition routes are summed, weighted by the inverse mismatch between the energy of the photons absorbed and the one-photon transition energies. 
The two-photon line strength factor $S\ri{ge}$ is given therefore as \cite{hippler99a}
\begin{align}
S\ri{ge} = \sum_{\Me, \Mg} \Bigg | \sum_{\mathrm i} \frac{1}{\oig - \omega_1}\Braket{\mathrm{e}  | \esigma \cdot \boldsymbol{\myhat\mu} | \mathrm{i}}  \Braket{\mathrm{i} | \esigma \cdot \boldsymbol{\myhat\mu} | \mathrm{g}}\Bigg | ^{2}.
\label{eq_two_photon_LS_very_first_formula}
\end{align}
Here, $\ket{\mathrm{i}}$ is the intermediate state of the virtual one-photon transition route with the sum over $\mathrm i$ including all accessible intermediate states. $M\ri g$ and $M\ri e$ label the different Zeeman states in the ground and the excited state, respectively. The term $\oig - \omega_1$ represents the mismatch between the ground-intermediate-state transition energy $\hbar \oig$ and the photon energy $\hbar \omega_1$ (see Fig.~\ref{fig:two-photon_mechanism}~(a)). Moreover, $\boldsymbol{\myhat\mu} $ is the electric-dipole operator and $\esigma$ the unit polarization vector of the radiation with $\exPolSF = 0$ standing for linear, $\exPolSF = \pm 1$ for circular polarization.

Assuming that the molecular states may be written as a product of an electronic-vibrational state (labeled ``ev'') and a nuclear-spin-rotational state (labeled ``nsr''), the two-photon line strength in Eq.~\eqref{eq_two_photon_LS_very_first_formula} takes the form:
\begin{align}
S\ri{ge}  =&  \sum_{\Me, \Mg} \Bigg | \sum_{\mathrm{i}_\mathrm{ev}} \sum_{\mathrm{i}_\mathrm{nsr}} \frac{1}{\oievnsrg - \omega_1} 
\nonumber \\ 
& \times \Braket{\mathrm{e}\ri{ev},  \mathrm{e}\ri{nsr} | \esigma \cdot \boldsymbol{\myhat\mu} | \mathrm{i}\ri{ev},  \mathrm{i}\ri{nsr}} 
\nonumber \\ 
& \times \Braket{\mathrm{i}\ri{ev},  \mathrm{i}\ri{nsr} | \esigma \cdot \boldsymbol{\myhat\mu} | \mathrm{g}\ri{ev}, \mathrm{g}\ri{nsr}}\Bigg | ^{2}.
\label{eq_two_photon_LS_very_2nd_formula}
\end{align}
Here the sum over all intermediate states has been written as a sum over all electronic-vibrational intermediate states $\sum_{\mathrm{i}_\mathrm{ev}}$ and all nuclear-spin-rotational intermediate states $\sum_{\mathrm{i}\ri{nsr}}$. Accordingly, $\oig$ has been rewritten as $\oievnsrg$.

To evaluate Eq.~\eqref{eq_two_photon_LS_very_2nd_formula}, we express the scalar product $\esigma \cdot \boldsymbol{\myhat\mu}$ in spherical tensor notation, change to the molecule-fixed frame by the aid of Wigner rotation matrices and  factorize the transition matrix elements into purely angular and purely vibronic terms. For the ground-intermediate matrix element we get,
\begin{align}
&\Braket{ \mathrm{e}\ri{ev},  \mathrm{e}\ri{nsr} | \esigma \cdot \boldsymbol{\myhat\mu} | \mathrm{i}\ri{ev},  \mathrm{i}\ri{nsr}} \nonumber \\
& =\sum_{\exPolMFone =-1}^{1} \Braket{ \mathrm{e}\ri{ev},  \mathrm{e}\ri{nsr} | \left [\WignerDop{(1)}{\exPolSF}{\exPolMFone} \right]^{*} \spht{1}{\exPolMFone}{\boldsymbol{\myhat\mu} }| \mathrm{i}\ri{ev},  \mathrm{i}\ri{nsr}} \\ 
&= \sum_{\exPolMFone =-1}^{1}\Braket{ \mathrm{e}\ri{ev} |  \spht{1}{\exPolMFone}{\boldsymbol{\myhat\mu} }| \mathrm{i}\ri{ev}
} \Braket{  \mathrm{e}\ri{nsr}  |  \left[ \WignerDop{(1)}{\exPolSF}{\exPolMFone} \right ]^{*}  | \mathrm{i}\ri{nsr}} 
.
\end{align}
For the intermediate-excited matrix element, an analogous expression is obtained.

Thus, the two-photon line strength is:
\begin{align}
S\ri{ge} = & \sum_{\Me, \Mg} \Bigg | \sum_{\mathrm{i}_\mathrm{ev}} \sum_{\mathrm{i}_\mathrm{nsr}} \frac{1}{\oievnsrg - \omega_1}\sum_{\exPolMFone,\exPolMFtwo} \Braket{ \mathrm{e}\ri{ev} | \spht{1}{\exPolMFone}{\boldsymbol{\myhat\mu} } |  \mathrm{i}\ri{ev} } \nonumber \\
& \times \Braket{  \mathrm{i}\ri{ev} | \spht{1}{\exPolMFtwo}{\boldsymbol{\myhat\mu} } | \mathrm{g}\ri{ev}}
\Braket{  \mathrm{e}\ri{nsr}  |  \left[ \WignerDop{(1)}{\exPolSF}{\exPolMFone} \right ]^{*}  |  \mathrm{i}\ri{nsr} } \nonumber \\ 
&\times \Braket{  \mathrm{i}\ri{nsr} |  \left[ \WignerDop{(1)}{\exPolSF}{\exPolMFtwo} \right ]^{*}  |  \mathrm{g}\ri{nsr} }
\Bigg | ^{2}.
\end{align}
As indicated by the notation, the energy difference between ground and intermediate level $\hbar\oievnsrg$ depends in principle on both, the vibronic and the nuclear-spin-rotational state of the intermediate level. However, since the nuclear-spin-rotational contribution to $\oievnsrg$ is small compared to the vibronic one, we may neglect the latter and approximate the energy mismatch for far-off-resonant excitation\footnote{Although the overall [2+1'] REMPI scheme is a resonant process (in the sense that ionization is achieved via a neutral excited state), the two-photon, bound-bound transition connecting the neutral ground state to the neutral excited state regarded by itself is a \emph{non}-resonant process.} as $ \oievnsrg - \omega_1 \approx  \omega_{\mathrm{i}\ri{ev}\mathrm{g}} - \omega_1$. Doing so, the term $1/(\oievnsrg - \omega_1) \approx  1/(\omega_{\mathrm{i}\ri{ev}\mathrm{g}} - \omega_1)$ may be factored out of the sum over the nuclear-spin-rotational intermediate states ($\sum_{\mathrm{i}\ri{nsr}}$) and the expression for the line strength separates into a product of two independent sums, $\sum_{\mathrm{i}\ri{ev}}$ and $\sum_{\mathrm{i}\ri{nsr}}$:
\begin{align}
S\ri{ge} = & \sum_{\Me, \Mg} \Bigg |  \sum_{\exPolMFone, \exPolMFtwo} \bigg (
\sum_{\mathrm{i}_\mathrm{ev}} \frac{1}{\omega_{\mathrm{i}\ri{ev}\mathrm{g}}- \omega_1}
\Braket{ \mathrm{e}\ri{ev} | \spht{1}{\exPolMFone}{\boldsymbol{\myhat\mu} } | \mathrm{i}\ri{ev} } \nonumber \\
& \times \Braket{  \mathrm{i}\ri{ev} | \spht{1}{\exPolMFtwo}{\boldsymbol{\myhat\mu} } | \mathrm{g}\ri{ev} }
\bigg ) \bigg ( \sum_{\mathrm{i}\ri{nsr}} \Braket{  \mathrm{e}\ri{nsr}  |  \left[ \WignerDop{(1)}{\exPolSF}{\exPolMFone} \right ]^{*}  | \mathrm{i}\ri{nsr} } \nonumber \\
& \times \Braket{ \mathrm{i}\ri{nsr}  |  \left[ \WignerDop{(1)}{\exPolSF}{\exPolMFtwo} \right ]^{*}  |  \mathrm{g}\ri{nsr} } \bigg )
\Bigg | ^{2}.
\end{align}

Without the energy-mismatch weighting factors, the sum over the intermediate nuclear-spin-rotational states is a sum of projection operators $\Ket{ \mathrm{i}\ri{nsr} } \Bra{ \mathrm{i}\ri{nsr}}$. Since this sum includes \emph{all} nuclear-spin-rotational states, it is equal to the identity operator $\myhat{  I}\ri{nsr}$ for the nuclear-spin-rotational states, i.e.,
\begin{equation}
\sum_{\mathrm{i}\ri{nsr}} \Ket{ \mathrm{i}\ri{nsr} } \Bra{ \mathrm{i}\ri{nsr} } = \myhat{ I}\ri{nsr}.
\end{equation}
Therefore, we arrive at
\begin{align}
S\ri{ge} = &\sum_{\Me, \Mg} \Bigg |  \sum_{\exPolMFone,\exPolMFtwo}  \Braket{  \mathrm{e}\ri{nsr}  |  \left[ \WignerDop{(1)}{\exPolSF}{\exPolMFone} \right ]^{*}  \left[ \WignerDop{(1)}{\exPolSF}{\exPolMFtwo} \right ]^{*}  |  \mathrm{g}\ri{nsr} } \nonumber \\  
& \times \sum_{\mathrm{i}_\mathrm{ev}} \frac{1}{\omega_{\mathrm{i}\ri{ev}\mathrm{g}}- \omega_1} \Braket{ \mathrm{e}\ri{ev} | \spht{1}{\exPolMFone}{\boldsymbol{\myhat\mu} } | \mathrm{i}\ri{ev}} \nonumber \\
& \times \Braket{ \mathrm{i}\ri{ev} | \spht{1}{\exPolMFtwo}{\boldsymbol{\myhat\mu} } | \mathrm{g}\ri{ev}} \Bigg | ^{2} .
\end{align}

To proceed, we need to choose a basis for the molecular states. We focus on the frequent case of transitions between $^{1}\Sigma$ states (and note that the following treatment can be adapted to other state symmetries and coupling cases, see Refs.~\onlinecite{hippler99a,lefebvre-brion04a}). We chose the Hund's case (b) notation for the electronic-vibrational ground, intermediate and excited states:
\begin{subequations}
\begin{align}
&\Ket{ \mathrm{g}\ri{ev} } =  \Ket{n'' \Lambda'', v''},
\\ &\Ket{  \mathrm{i}\ri{ev} }  =  \Ket{n_{\mathrm{i}\ri{ev}}  \Lambda_{\mathrm{i}\ri{ev}}, v_{\mathrm{i}\ri{ev}}},
\\ &\Ket{  \mathrm{e}\ri{ev} } =  \Ket{n' \Lambda', v'}.
\end{align}
\end{subequations}
Here, $n''$, $n_{\mathrm{i}\ri{ev}}$, $n'$ denote the electronic ground, intermediate and excited level, respectively. $v''$, $v_{\mathrm{i}\ri{ev}}$, $v'$ stand for the corresponding vibrational states and $\Lambda''$, $ \Lambda_{\mathrm{i}\ri{ev}}$, $\Lambda'$ are the projections of the total electron orbital angular momenta on the internuclear axis. Refer to Tab.~\ref{tab:REMPI_nomenclature} for a summary of the symbols used in the present work.

The angular part of the ground and excited state are written as
\begin{subequations}
\begin{align}
&\Ket{ \mathrm{g}\ri{nsr} } =  \Ket{N'' \Lambda'' I'' F'' M_{F}''},
\\ &\Ket{  \mathrm{e}\ri{nsr} } =  \Ket{N' \Lambda' I' F' M_{F}'},
\end{align}
\end{subequations}
with $N''$ and $N'$ the rotational quantum numbers in the ground and the excited state, $I''$ and $I'$ the respective nuclear spin quantum numbers, $F''$ and $F'$ the total angular momentum quantum numbers as well as $M_{F}''$, $ M_{F}'$ the corresponding  angular momentum projection quantum numbers. 

\begin{table*}[tpb]
\centering
\caption{Symbols used in the model of the [2+1'] REMPI process}
\label{tab:REMPI_nomenclature}
\newlength{\symsREMPITableSpace}
\setlength{\symsREMPITableSpace}{0.7mm}
\begin{footnotesize}
\begin{tabular}{l l l l}
\shortstack[l]{Quantum number\\(magnitude)}	& \shortstack[l]{Mol.-fixed\\projection} 	&	\shortstack[l]{Space-fixed\\projection} 	&	Description\\
\hline
\rule{0mm}{3.5mm}
$n'' $		&	-			&	-			&	Label for the electronic state in the neutral ground state (AB)\\[\symsREMPITableSpace]
$ v'' $	&	-			&	-			&	Vibrational quantum number in the neutral ground state\\[\symsREMPITableSpace]

$ N'' $	 & 	$ \Lambda'' $	&	$ M_N'' $		&	Orbital-rotational angular momentum in the neutral ground state \\[\symsREMPITableSpace]
$ I'' $		 & 	-			&	$ M_I'' $		&	Nuclear spin in the neutral ground state  \\[\symsREMPITableSpace]
$ F'' $	 & 	-			&	$ M_F'' $		&	Total angular momentum in the neutral ground state  \\[\symsREMPITableSpace]

$ n_{\mathrm{i}\ri{ev}}  $	&	-			&	-			&	Label for the electronic state in the intermediate state of the \\
														&&&two-photon transition of the excitation step ($\text{AB} \rightarrow \text{AB*}$) \\[\symsREMPITableSpace]
$ v_{\mathrm{i}\ri{ev}}  $	&	-			&	-			&	Vibrational quantum number in the intermediate state of the\\
														&&&two-photon transition in the excitation step \\[\symsREMPITableSpace]
$ n' $	&	-			&	-			&	Label for the electronic state in the neutral, excited state (AB*)\\[\symsREMPITableSpace]
$ v' $		&	-			&	-			&	Vibrational quantum number in the neutral, excited state \\[\symsREMPITableSpace]
$ N' $	 &	$ \Lambda' $	&	 $ M_N' $		&	Orbital-rotational angular momentum in the neutral, excited state \\[\symsREMPITableSpace]
$ I' $	 	& 	-			&	$ M_I' $		&	Nuclear spin in the neutral, excited state \\[\symsREMPITableSpace]
$ F' $	 & 	-			&	$ M_F' $		&	Total angular momentum in the neutral, excited state \\[\symsREMPITableSpace]

$\np $		&	-			&	-			&	Label for the electronic state of the molecular ion (AB$^+$)\\[\symsREMPITableSpace]
$ \vp $	&	-			&	-			&	Vibrational quantum number of the molecular ion\\[\symsREMPITableSpace]
$ \Np $	 & 	$ \Lp $		&	$ \MNp $		&	Orbital-rotational angular momentum of the molecular ion \\[\symsREMPITableSpace]
$ \Sp $	 & 	-			&	$ \MSp $		&	Electron spin of the molecular ion  \\[\symsREMPITableSpace]
$ \Jp $	 & 	-			&	$ \MJp $		&	Total angular momentum of the molecular ion excluding nuclear spin  \\[\symsREMPITableSpace]
$ \Ip $	 & 	-			&	$ \MIp $		&	Nuclear spin of the molecular ion  \\[\symsREMPITableSpace]
$ \Fp $	 & 	-			&	$ \MFp $		&	Total angular momentum of the molecular ion  \\[\symsREMPITableSpace]
1		&$\exPolMFone$ ($= \exPolMFoneNoIndex$)	&	$ \exPolSF $	&	Angular momentum of the first photon in the excitation step \\[\symsREMPITableSpace]
1		&$\exPolMFtwo$	&	$ \exPolSF $	&	Angular momentum of the second photon in the excitation step \\[\symsREMPITableSpace]
$\kEx$&$\exPolMFone + \exPolMFtwo$&	$2\exPolSF$&	Total angular momentum transferred to/from the molecule in the \\
										&&&	excitation step \\[\symsREMPITableSpace]
1	 	&	-			&	$\mu_0$ 		&	Angular momentum of the photon in the ionization step ($\text{AB*} \rightarrow \text{AB}^+$) \\[\symsREMPITableSpace]

$ l $		 &	-			&	$ m_l $		&	Orbital angular momentum of the photoelectron \\[\symsREMPITableSpace]
$ s $		 & 	-			&	$ m_s $		&	Spin of the photoelectron ($s=1/2$)       \\[\symsREMPITableSpace]

$ k $	 	& 	$q$			&	$p$			&	Total orbital angular momentum transferred to/from the molecule \\
										&&&	in the ionization step ($ p=-m_l+\mu_0 $) \\[\symsREMPITableSpace]
$ u $	 	& 	-			&	$\upr$		&	Total angular momentum transferred to/from the molecule in the\\
										&&& ionization step ($ \upr = -m_s + p $) \\[\symsREMPITableSpace]
\hline
\end{tabular}
\end{footnotesize}
\end{table*}

As we are assuming $\Sigma$ states for the ground and the excited state, we have $\Lambda'' = \Lambda' = 0$. For the intermediate state, also states with $ \Lambda_{\mathrm{i}\ri{ev}} \neq 0$ need to be considered (see Fig.~\ref{fig:two-photon_mechanism}~(b)).

Using this notation, the line strength for the two-photon transition is:
\begin{widetext}
\begin{align}
\SppTop = 
\sum_{M'_{F}, M''_{F}} \Bigg | 
 &
  \sum_{\exPolMFone,\exPolMFtwo} 
 \Braket{N' \Lambda' I' F' M_{F}'  |  \left[ \WignerDop{(1)}{\exPolSF}{\exPolMFone} \right ]^{*} 
 \left[ \WignerDop{(1)}{\exPolSF}{\exPolMFtwo} \right ]^{*}  |  N'' \Lambda'' I'' F'' M_{F}'' } 
\nonumber  \\ &\sum_{\mathrm{i}_\mathrm{ev}} \frac{1}{\omega_{\mathrm{i}\ri{ev}\mathrm{g}} - \omega_1}
\Braket{n' \Lambda', v' | \spht{1}{\exPolMFone}{\boldsymbol{\myhat\mu} } | n_{\mathrm{i}\ri{ev}}  \Lambda_{\mathrm{i}\ri{ev}}, v_{\mathrm{i}\ri{ev}}}
\Braket{ n_{\mathrm{i}\ri{ev}}  \Lambda_{\mathrm{i}\ri{ev}}, v_{\mathrm{i}\ri{ev}} | \spht{1}{\exPolMFtwo}{\boldsymbol{\myhat\mu} } |n'' \Lambda'', v''} 
\Bigg | ^{2} .
\label{eq_two_photon_line_strength_with_basis}
\end{align}
\end{widetext}

Exploiting that the nuclear-spin states are not affected in electric-dipole transitions, we decouple the nuclear spin from the total angular momentum in the ground state according to
\begin{multline}
\Ket{N'' \Lambda'' I'' F'' M_{F}''} = \\
\sum_{M''_N, M''_I} \CG{N''}{M''_N}{I''}{M''_I}{F''}{M''_F} \Ket{N'' \Lambda'' M''_N, I'' M_{I}''},
\end{multline}
with the Clebsch-Gordan coefficients $\CG{N''}{M''_N}{I''}{M''_I}{F''}{M''_F}$, and analogously in the excited state.

The angular matrix element in Eq.~\eqref{eq_two_photon_line_strength_with_basis} thus accounts for:
\begin{align}
& \Braket{N' \Lambda' I' F' M_{F}'  |  \left[ \WignerDop{(1)}{\exPolSF}{\exPolMFone} \right ]^{*}  \left[ \WignerDop{(1)}{\exPolSF}{\exPolMFtwo} \right ]^{*}  |  N'' \Lambda'' I'' F'' M_{F}'' } 
 \nonumber \\
& \begin{aligned}
 %
%
%
%
  \\ = \: & \kd{I'}{I''} (-1)^{N'-I''+M'_F} (-1)^{N''-I''+M''_F} \sqrt{2 F'+1} \sqrt{2 F''+1} \\
 & \times \sum_{M'_N, M''_N} \Braket{N' \Lambda' M'_N|  \left[ \WignerDop{(1)}{\exPolSF}{\exPolMFone} \right ]^{*}  \left[ \WignerDop{(1)}{\exPolSF}{\exPolMFtwo} \right ]^{*}  | N'' \Lambda'' M''_N}  \\
  & \times    \sum_{M''_I} \tj{N'}{I''}{F'}{M'_N}{M''_I}{-M'_F}  \tj{N''}{I''}{F''}{M''_N}{M''_I}{-M''_F} 
  , 
\label{eq_two_photon_angular_mat_el_with_two_WignerD}
 \end{aligned}
\end{align}
where the orthonormality of the nuclear spin states has been used and the Clebsch-Gordan coefficients have been replaced by 3j-symbols.

The rotational matrix element on the next-to-last line in Eq.~\eqref{eq_two_photon_angular_mat_el_with_two_WignerD} can be reformulated
using the relation\cite{zare88a, edmonds64a}
\begin{align}
\WignerDop{(j_{1})}{m'_{1}}{m_{1}} \WignerDop{(j_{2})}{m'_{2}}{m_{2}} 
= &
\sum_{j_{3} = |j_{1} - j_{2} |}^{j_{1} + j_{2}}
(2 j_{3} +1) \tj{j_1}{j_2}{j_3}{m'_1}{m'_2}{m'_3}
\nonumber \\ & \times
 \tj{j_1}{j_2}{j_3}{m_1}{m_2}{m_3}
[ \WignerDop{(j_{3})}{m'_{3}}{m_{3}} ]^{*}
, 
\end{align}
as \cite{hippler99a,germann16a}
\begin{align}
&\Braket{N' \Lambda' M'_N|  \left[ \WignerDop{(1)}{\exPolSF}{\exPolMFone} \right ]^{*}  \left[ \WignerDop{(1)}{\exPolSF}{\exPolMFtwo} \right ]^{*}  | N'' \Lambda'' M''_N} \nonumber \\
& = \sum_{\kEx=0}^{2} (2\kEx+1) \tj{1}{1}{\kEx}{-\exPolSF}{-\exPolSF}{2\exPolSF}  \tj{1}{1}{\kEx}{-\exPolMFtwo}{-\exPolMFone}{\exPolMFone+\exPolMFtwo} \nonumber \\ 
& \quad \times \Braket{N' \Lambda' M'_N | \WignerDop{(\kEx)}{-2\exPolSF, \;}{-\exPolMFone-\exPolMFtwo} | N'' \Lambda'' M''_N}.
\label{eq_two_photon_rotational_mat_el}
\end{align}

Inserting appropriately normalized Wigner rotation matrices for the rotational states (with the three Euler angles $\phi$, $\theta$, $\chi $),
\begin{align}
\braket{ \phi \, \theta \, \chi | N \Lambda M_N} &=  \sqrt{\frac{2N+1}{8\pi^2}} \left[ \WignerDop{(N)}{M_N}{\Lambda}( \phi, \theta, \chi) \right]^{*}
,
\end{align}
we obtain for the matrix element in  Eq.~\eqref{eq_two_photon_rotational_mat_el} an integral over three Wigner rotation matrices that may be expressed in the form of 3j-symbols as \cite{edmonds64a,zare88a}
\begin{align}
&\Braket{N' \Lambda' M'_N | \WignerDop{(\kEx)}{-2\exPolSF, \;}{-\exPolMFone-\exPolMFtwo} | N'' \Lambda'' M''_N}
\nonumber \\
 &= \sqrt{2N'+1}  \sqrt{2N''+1} (-1)^{M''_N - \Lambda''} \tj{N'}{\kEx}{N''}{M'_N}{-2 \exPolSF}{-M''_N} \nonumber \\
 & \quad \times \tj{N'}{\kEx}{N''}{\Lambda'}{- \exPolMFone - \exPolMFtwo}{-\Lambda''}.
\end{align}

Substituting this expression into Eq.~\eqref{eq_two_photon_rotational_mat_el} yields,
\begin{align}
&\Braket{N' \Lambda' M'_N|  \left[ \WignerDop{(1)}{\exPolSF}{\exPolMFone} \right ]^{*}  \left[ \WignerDop{(1)}{\exPolSF}{\exPolMFtwo} \right ]^{*}  | N'' \Lambda'' M''_N} \nonumber \\ 
&=\sqrt{2N'+1}  \sqrt{2N''+1}  (-1)^{M''_N - \Lambda''} \nonumber \\
& \quad \times \sum_{\kEx=0}^{2} (2\kEx+1) \tj{1}{1}{\kEx}{-\exPolSF}{-\exPolSF}{2\exPolSF}  \tj{1}{1}{\kEx}{-\exPolMFtwo}{-\exPolMFone}{\exPolMFone+\exPolMFtwo} \nonumber \\ 
& \quad \times \tj{N'}{\kEx}{N''}{M'_N}{-2 \exPolSF}{-M''_N} \tj{N'}{\kEx}{N''}{\Lambda'}{- \exPolMFone - \exPolMFtwo}{-\Lambda''},
\end{align}
and subsequent substitution into Eq.~\eqref{eq_two_photon_angular_mat_el_with_two_WignerD} gives
\begin{align}
& \Braket{N' \Lambda' I' F' M_{F}'  |  \left[ \WignerDop{(1)}{\exPolSF}{\exPolMFone} \right ]^{*}  \left[ \WignerDop{(1)}{\exPolSF}{\exPolMFtwo} \right ]^{*}  |  N'' \Lambda'' I'' F'' M_{F}'' } \nonumber \\ 
& = \kd{I'}{I''} (-1)^{N'-I''+M'_F} (-1)^{N''-I''+M''_F} (-1)^{-\Lambda''} \nonumber \\
& \quad\times \sqrt{2 F'+1} \sqrt{2 F''+1}  \sqrt{2 N'+1} \sqrt{2 N''+1}
 \nonumber \\ 
 &\quad \times \sum_{\kEx=0}^{2} (2\kEx+1) \tj{1}{1}{\kEx}{-\exPolSF}{-\exPolSF}{2\exPolSF}  \tj{1}{1}{\kEx}{-\exPolMFtwo}{-\exPolMFone}{\exPolMFone+\exPolMFtwo} \nonumber \\
 & \quad \times \tj{N'}{\kEx}{N''}{\Lambda'}{- \exPolMFone - \exPolMFtwo}{-\Lambda''} \sum_{M'_N, M''_N, M''_I} (-1)^{M''_N}
  \nonumber \\
 & \quad \times  \tj{N'}{\kEx}{N''}{M'_N}{-2 \exPolSF}{-M''_N}\tj{N'}{I''}{F'}{M'_N}{M''_I}{-M'_F} \nonumber \\
 &\quad \times  \tj{N''}{I''}{F''}{M''_N}{M''_I}{-M''_F}.
\label{eq_two_photon_angular_mat_el_not_yet_simplified}
\end{align}

The sum over the last three 3j-symbols in Eq.~\eqref{eq_two_photon_angular_mat_el_not_yet_simplified} may be expressed in terms of a Wigner 6j-symbol \cite{zare88a} yielding:
\begin{multline}
 \Braket{N' \Lambda' I' F' M_{F}'  |  \left[ \WignerDop{(1)}{\exPolSF}{\exPolMFone} \right ]^{*}  \left[ \WignerDop{(1)}{\exPolSF}{\exPolMFtwo} \right ]^{*}  |  N'' \Lambda'' I'' F'' M_{F}'' } 
\\
\begin{aligned} 
 =  \: & \kd{I'}{I''} (-1)^{N' + N'' -\Lambda'' - 3I'' -F' -F'' +M'_F + 2 M''_F}
 \\ &
 \sqrt{2 F'+1} \sqrt{2 F''+1}  \sqrt{2 N'+1} \sqrt{2 N''+1}
\\ & \sum_{\kEx=0}^{2} (2\kEx+1) \tj{1}{1}{\kEx}{-\exPolSF}{-\exPolSF}{2\exPolSF}  \tj{1}{1}{\kEx}{-\exPolMFtwo}{-\exPolMFone}{\exPolMFone+\exPolMFtwo} 
\\ & \tj{N'}{\kEx}{N''}{\Lambda'}{- \exPolMFone - \exPolMFtwo}{-\Lambda''} \sj{\kEx}{N''}{N'}{I''}{F'}{F''} 
\\& \tj{F'}{\kEx}{F''}{-M'_F}{2\exPolSF}{M''_F} .
\end{aligned}
\label{eq_two_photon_term_with_6j_symbol}
\end{multline}
Because of  the third 3j-symbol in Eq.~\eqref{eq_two_photon_term_with_6j_symbol}, this matrix element vanishes unless the condition $\exPolMFone +\exPolMFtwo = \Lambda' - \Lambda''$ ($=0$ for $\Sigma$-$\Sigma$ transitions) is met. Hence, when substituting this matrix element into Eq.~\eqref{eq_two_photon_line_strength_with_basis}, only terms fulfilling this relation contribute to the sums over $\exPolMFone$ and $\exPolMFtwo$.  We thus skip the sum over $\exPolMFtwo$ in Eq.~\eqref{eq_two_photon_line_strength_with_basis} by means of the substitution $\exPolMFtwo = -\exPolMFone$ and drop the index on $\exPolMFone$ by setting $\exPolMFoneNoIndex \coloneqq \exPolMFone$. Furthermore, the cross terms in the sum over $\kEx$ in Eq.~\eqref{eq_two_photon_line_strength_with_basis} vanish when summing over $M'_F$ and $M''_F$ owing to the orthogonality properties of the last 3j-symbol in Eq.~\eqref{eq_two_photon_term_with_6j_symbol}. As a result we obtain the two-photon line strength as
\begin{align}
\SppTop = \:
& (2 F'+1) (2 F''+1) (2 N'+1) (2 N''+1)  \kd{I'}{I''} \nonumber \\
& \sum_{\kEx=0,2} (2\kEx+1) \tj{1}{1}{\kEx}{-\exPolSF}{-\exPolSF}{2\exPolSF}^2 
\nonumber \\ 
& \sj{\kEx}{N''}{N'}{I''}{F'}{F''}^2  
 \tj{N'}{\kEx}{N''}{0}{0}{0}^2
 U(\kEx),
\label{eq_two_photon_line_strength_general_result_1}
\end{align}
with 
\begin{align}
U(\kEx) = &
\Bigg | \sum_{\exPolMFoneNoIndex=-1}^{1}  \tj{1}{1}{\kEx}{\exPolMFoneNoIndex}{-\exPolMFoneNoIndex}{0}
      \sum_{\mathrm{i}_\mathrm{ev}} \frac{1}{\omega_{\mathrm{i}\ri{ev}\mathrm{g}} - \omega_1} \nonumber \\
& \times \Braket{n' \: \Lambda'=0, v' | \spht{1}{\exPolMFoneNoIndex}{\boldsymbol{\myhat\mu} } | n_{\mathrm{i}\ri{ev}}  \Lambda_{\mathrm{i}\ri{ev}}, v_{\mathrm{i}\ri{ev}}} \nonumber \\ &
\times \Braket{ n_{\mathrm{i}\ri{ev}}  \Lambda_{\mathrm{i}\ri{ev}}, v_{\mathrm{i}\ri{ev}} | \spht{1}{-\exPolMFoneNoIndex}{\boldsymbol{\myhat\mu} } |n'' \:\Lambda''=0, v''} 
\Bigg | ^{2} .
\label{eq_U-term}
\end{align}

Since the first 3j-symbol in  Eq.~\eqref{eq_two_photon_line_strength_general_result_1} vanishes for $\kEx = 1$, we have omitted the $\kEx=1$-term in this equation. We study the expression $U(\kEx)$ separately for $\kEx=0$ and $\kEx=2$. For $\kEx=0$, evaluation of the 3j-symbol in Eq.~\eqref{eq_U-term}  yields
\begin{align}
U(0) 
= 
\: &\frac{1}{3} \left | \mu_{||}\mu'_{||} - \mu_{+}\mu'_{-} - \mu''_{-}\mu''_{+} \right |^2
=
\frac{1}{3} \mu_\mathrm{I}^{2},
\end{align}
with the abbreviations (see Refs.~\onlinecite{bray76a,hanisco91a})\footnote{For the signs of $\mu_{+}\mu'_{-}$ and $\mu''_{-}\mu''_{+}$ different conventions are found in the literature. Here, the same sign convention as in Ref.~\onlinecite{hanisco91a} has been chosen, differing from the one used in Ref.~\onlinecite{bray76a}.}
\begin{subequations}
\begin{align}
\mu_{||} \mu'_{||} &= \sum_{\mathrm{i}_\mathrm{ev}} \frac{1}{\omega_{\mathrm{i}\ri{ev}\mathrm{g}} - \omega_1}
\Braket{n' \Lambda', v' | \spht{1}{0}{\boldsymbol{\myhat\mu} } | n_{\mathrm{i}\ri{ev}}  \Lambda_{\mathrm{i}\ri{ev}}, v_{\mathrm{i}\ri{ev}}} \nonumber \\
& \quad \times \Braket{ n_{\mathrm{i}\ri{ev}}  \Lambda_{\mathrm{i}\ri{ev}}, v_{\mathrm{i}\ri{ev}} | \spht{1}{0}{\boldsymbol{\myhat\mu} } |n'' \Lambda'', v''} , \\
\mu_{+}\mu'_{-} &= \sum_{\mathrm{i}_\mathrm{ev}} \frac{1}{\omega_{\mathrm{i}\ri{ev}\mathrm{g}} - \omega_1}
\Braket{n' \Lambda', v' | \spht{1}{+1}{\boldsymbol{\myhat\mu} } | n_{\mathrm{i}\ri{ev}}  \Lambda_{\mathrm{i}\ri{ev}}, v_{\mathrm{i}\ri{ev}}} \nonumber \\
& \quad \times \Braket{ n_{\mathrm{i}\ri{ev}}  \Lambda_{\mathrm{i}\ri{ev}}, v_{\mathrm{i}\ri{ev}} | \spht{1}{-1}{\boldsymbol{\myhat\mu} } |n'' \Lambda'', v''} , \\
\mu''_{-}\mu''_{+} &= \sum_{\mathrm{i}_\mathrm{ev}} \frac{1}{\omega_{\mathrm{i}\ri{ev}\mathrm{g}} - \omega_1}
\Braket{n' \Lambda', v' | \spht{1}{-1}{\boldsymbol{\myhat\mu} } | n_{\mathrm{i}\ri{ev}}  \Lambda_{\mathrm{i}\ri{ev}}, v_{\mathrm{i}\ri{ev}}} \nonumber \\
& \quad \times \Braket{ n_{\mathrm{i}\ri{ev}}  \Lambda_{\mathrm{i}\ri{ev}}, v_{\mathrm{i}\ri{ev}} | \spht{1}{+1}{\boldsymbol{\myhat\mu} } |n'' \Lambda'', v''} ,
\end{align}
\label{eq_two-photon_dipole_moments}
\end{subequations}
and
\begin{equation}
\mu_\mathrm{I}^{2} = \left | \mu_{||}\mu'_{||} - \mu_{+}\mu'_{-} - \mu''_{-}\mu''_{+} \right |^2.
\end{equation}
For $\kEx=2$, we obtain similarly
\begin{equation}
U(2) = \frac{1}{30}   \left | 2 \mu_{||}\mu'_{||} + \mu_{+}\mu'_{-} + \mu''_{-}\mu''_{+} \right |^2
=  \frac{1}{30} \mu_\mathrm{S}^2,
\label{eq_U(k=2)}
\end{equation}
where we have set $\mu_\mathrm{S}^2 = \left | 2 \mu_{||}\mu'_{||} + \mu_{+}\mu'_{-} + \mu''_{-}\mu''_{+} \right |^2$.

The hfs-resolved two-photon line strength between $^{1}\Sigma$-states is thus:
\begin{align}
\SppTop = \:
& (2 F'+1) (2 F''+1) (2 N'+1) (2 N''+1)  \kd{I'}{I''} \nonumber \\ 
& \times \left [  \frac{1}{3}  \frac{1}{2N''+1} \tj{1}{1}{0}{-\exPolSF}{-\exPolSF}{2\exPolSF}^2 \sj{0}{N''}{N'}{I''}{F'}{F''}^2 \mu_\mathrm{I}^{2} \right .
\nonumber \\ 
& + \frac{1}{6} \tj{N'}{2}{N''}{0}{0}{0}^2 \tj{1}{1}{2}{-\exPolSF}{-\exPolSF}{2\exPolSF}^2 \nonumber \\
& \times \sj{2}{N''}{N'}{I''}{F'}{F''}^2 \mu_\mathrm{S}^2 \Bigg ].
\label{eq_two_photon_line_strength_general_result_2}
\end{align}

For circular polarized radiation, we have $\exPolSF = \pm 1$. As $\exPolSF$ denotes the projection associated with $\kEx$ on the space-fixed $z$-axis, we must have $\kEx \ge \exPolSF$. Hence, the term with $\kEx=0$ in Eq.~\eqref{eq_two_photon_line_strength_general_result_1}, i.e., the first summand within brackets in Eq.~\eqref{eq_two_photon_line_strength_general_result_2}, does not apply for circular polarization. With the last 3j-symbol
in Eq.~\eqref{eq_two_photon_line_strength_general_result_2}
accounting for 1/5, we thus obtain:
\begin{align}
S_{'' \leftrightarrow '}^\text{(circ)}  = \:
&   \frac{1}{30} (2 F'+1) (2 F''+1) (2 N'+1) (2 N''+1)  \kd{I'}{I''} \nonumber \\
& \times \tj{N'}{2}{N''}{0}{0}{0}^2 \sj{2}{N''}{N'}{I''}{F'}{F''}^2 \mu_\mathrm{S}^2.
\end{align}

For linear polarization, we have $\exPolSF=0$ for a suitable chosen space-fixed frame of reference. The two squared 3j-symbols involving $\exPolSF$ in Eq.~\eqref{eq_two_photon_line_strength_general_result_2} then account for 1/3 and 2/15, respectively, and the line strength is 
\begin{align}
S_{'' \leftrightarrow '}^\text{(lin)} = \:
& (2 F'+1) (2 F''+1) (2 N'+1) (2 N''+1)  \kd{I'}{I''} \nonumber \\ 
& \times \Bigg [  \frac{1}{9}  \frac{1}{2N''+1} \sj{0}{N''}{N'}{I''}{F'}{F''}^2 \mu_\mathrm{I}^{2} 
   \nonumber \\
 & + \frac{1}{45} \tj{N'}{2}{N''}{0}{0}{0}^2 \sj{2}{N''}{N'}{I''}{F'}{F''}^2 \mu_\mathrm{S}^2  \Bigg ].
\end{align}



Summing the above expression over all hyperfine components corresponding to a specific rotational transition reproduces the results for the rotationally resolved two-photon line strength reported in Ref.~\onlinecite{bray76a}.
 

So far, the line strength associated with the \emph{total} population in a $(N',I',F')$ level has been considered. This is the quantity usually of interest for transitions between bound states. For our particular purpose, namely to describe the REMPI process, the population in a certain Zeeman state $\ket{ N',I',F',M'_F}$ of the neutral, electronically excited state is needed. The relevant quantity is thus:  \footnote{The quantity $S(M\ri e)$ defined in Eq.~\eqref{eq_LS_hfs_anisotrop} does not fully comply with the usual definition of a spectroscopic line strength, which involves sums over all degenerate states of the initial and the final level. $S(M\ri e)$ is rather just a quantity proportional to the excitation rate populating a certain $M\ri e$-Zeeman state and hence to the population in this state after a given excitation period. Nonetheless, the symbol $S$ is used for this quantity here as well.}
\begin{equation}
S\ri{ge}(M\ri e) =
\sum_{M \ri g} \left | \sum_{\mathrm{i}}
\frac{1}{ \oig - \omega_1}
\braket{\text{e}| \esigma \cdot \boldsymbol{\myhat\mu}  | \mathrm{i}} \braket{\mathrm{i} | \esigma \cdot \boldsymbol{\myhat\mu}  | \text{g}}
 \right | ^2.
\label{eq_LS_hfs_anisotrop}
\end{equation}
For hfs-resolved transitions, this quantity is written in our notation as
\begin{widetext}
\begin{align}
S(F'',F',M'_F) = &
\sum_{M''_F} \Bigg | 
 \sum_{\exPolMFone,\exPolMFtwo} 
 \Braket{N' \Lambda' I' F' M_{F}'  |  \left[ \WignerDop{(1)}{\exPolSF}{\exPolMFone} \right ]^{*} 
 \left[ \WignerDop{(1)}{\exPolSF}{\exPolMFtwo} \right ]^{*}  |  N'' \Lambda'' I'' F'' M_{F}'' } 
\nonumber  \\ &\sum_{\mathrm{i}_\mathrm{ev}} \frac{1}{\omega_{\mathrm{i}\ri{ev}\mathrm{g}} - \omega_1}
\Braket{n' \Lambda', v' | \spht{1}{\exPolMFone}{\boldsymbol{\myhat\mu} } | n_{\mathrm{i}\ri{ev}}  \Lambda_{\mathrm{i}\ri{ev}}, v_{\mathrm{i}\ri{ev}}}
\Braket{ n_{\mathrm{i}\ri{ev}}  \Lambda_{\mathrm{i}\ri{ev}}, v_{\mathrm{i}\ri{ev}} | \spht{1}{\exPolMFtwo}{\boldsymbol{\myhat\mu} } |n'' \Lambda'', v''} 
\Bigg | ^{2},
\label{eq_two_photon_line_strength_aniso_with_basis}
\end{align}
\end{widetext}
where, as before, the nuclear-spin-rotational contribution to the energy mismatch has been neglected.

Substituting the angular transition matrix element from Eq.~\eqref{eq_two_photon_term_with_6j_symbol}
and applying the above-mentioned restrictions and substitutions for $\exPolMFone$, $\exPolMFtwo$ yields (for $\Sigma$-$\Sigma$-transitions):
\begin{multline}
S(F'',F',M'_F) 
\\
\begin{aligned}
= \: &(2F'+1)(2F''+1)(2N'+1)(2N''+1)
 \kd{I'}{I''}
\\ & \times
 \sum_{M''_F} \Bigg | \sum_{\kEx=0, 2} (2\kEx+1) \tj{N'}{\kEx}{N''}{0}{0}{0}
 \tj{1}{1}{\kEx}{-\exPolSF}{-\exPolSF}{2\exPolSF}
 \\ & \times
 \sj{\kEx}{N''}{N'}{I''}{F'}{F''}  \tj{F'}{\kEx}{F''}{-M'_F}{2\exPolSF}{M''_F}  
  \\ & \times
 \sum_{\exPolMFoneNoIndex}  \tj{1}{1}{\kEx}{\exPolMFoneNoIndex}{-\exPolMFoneNoIndex}{0} 
 \sum_{\mathrm{i}_\mathrm{ev}} \frac{1}{\omega_{\mathrm{i}\ri{ev}\mathrm{g}} - \omega_1}
  \\ &\times
\Braket{n'  \: \Lambda'=0, v' | \spht{1}{\exPolMFoneNoIndex}{\boldsymbol{\myhat\mu} } | n_{\mathrm{i}\ri{ev}}  \Lambda_{\mathrm{i}\ri{ev}}, v_{\mathrm{i}\ri{ev}}}
\\ & \times
\Braket{ n_{\mathrm{i}\ri{ev}}  \Lambda_{\mathrm{i}\ri{ev}}, v_{\mathrm{i}\ri{ev}} | \spht{1}{-\exPolMFoneNoIndex}{\boldsymbol{\myhat\mu} } |n'' \: \Lambda''=0, v''} 
\Bigg | ^{2}.
\end{aligned}
\label{eq_two_photon_line_strength_aniso_simplified}
\end{multline}

In contrast to the total line strength considered before, the above expression does not include a sum over the excited-state projection angular momentum quantum number $M'_F$. As a consequence, the orthogonality of the 3j-symbols
may not be used to eliminate the cross terms, as was possible in the derivation of Eq.~\eqref{eq_two_photon_line_strength_general_result_1}.
Hence, angular terms may not be separated from vibronic ones. Therefore, calculation of relative nuclear-spin-rotational intensities is in general only possible when knowing both, the magnitude and the \emph{phase} of the vibronic transition matrix elements. If the phases are unknown, relative intensities may in general not be determined (see also Sec.~\ref{sec:Ionization_step} below).

For transitions involving a change in the rotational angular momentum, i.e., O and S lines ($\Delta N = -2$ and +2, respectively), however, only the $\kEx=2$ term in Eq.~\eqref{eq_two_photon_line_strength_aniso_simplified} is relevant. We may thus simplify the above result further arriving at:
\begin{multline}
S_\text{S, O}(F'',F',M'_F) = 
\\
\begin{aligned}
\: &\frac{5}{6} (2F'+1)(2F''+1)(2N'+1)(2N''+1)   
 \\ & \times
 \kd{I'}{I''} \tj{1}{1}{2}{-\exPolSF}{-\exPolSF}{2\exPolSF}^2 \sj{2}{N''}{N'}{I''}{F'}{F''}^2  
\\ & \times
\sum_{M''_F} \tj{F'}{2}{F''}{-M'_F}{2\exPolSF}{M''_F}^2
\tj{N'}{2}{N''}{0}{0}{0}^{2}
  \mu_\mathrm{S}^{2},
\end{aligned}
\label{eq_two_photon_line_strength_aniso_S_and_O-Lines}
\end{multline}
with $\mu_\mathrm{S}^{2}$ as in Eq.~\eqref{eq_U(k=2)}.

In the case of linear polarized radiation ($\exPolSF=0$), this results in
\begin{multline}
S^\text{(lin)}_\text{S, O}(F'',F',M'_F) = 
\\
\begin{aligned}
 &\frac{1}{9} (2F'+1)(2F''+1)(2N'+1)(2N''+1) 
 \\ & \times
\kd{I'}{I''} \sj{2}{N''}{N'}{I''}{F'}{F''}^2  \tj{F'}{2}{F''}{-M'_F}{0}{M'_F}^2  
\\ & \times
\tj{N'}{2}{N''}{0}{0}{0}^2
  \mu_\mathrm{S}^{2}.
\end{aligned}
\label{eq_two_photon_line_strength_aniso_S_and_O-Lines_lin_pol.}
\end{multline}


\section{Ionization step}
\label{sec:Ionization_step}

\subsection{Effects of anisotropic populations}

Having discussed the excitation step (AB~$\rightarrow$~AB*), we now turn to the ionization step (AB*~$\rightarrow$~AB$^{+}$). 

\begin{figure}[tbp]
\begin{center}
\includegraphics{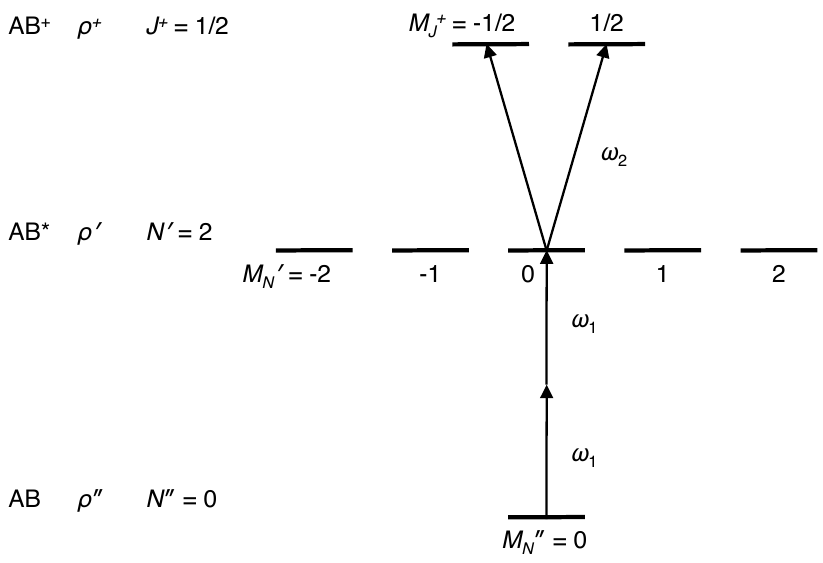}
\caption{{\bf Anisotropic population in the excited neutral state generated in REMPI:} example of the $N''=0 \rightarrow N'=2 \rightarrow \Np = 0$ REMPI scheme with linear polarization for excitation: because of the selection rule $M'_{N}=M''_{N}$ only the $M'_{N} = 0$ Zeeman state of the neutral excited level is populated. Ionization may thus only occur from this particular Zeeman state.}
\label{fig:Schema_anisotrope_Ionisation}
\end{center}
\end{figure}

The population of the neutral, excited molecules AB* produced in the excitation step is in general anisotropic, i.e., different Zeeman states are unequally populated.
This effect is illustrated in Fig.~\ref{fig:Schema_anisotrope_Ionisation} with hyperfine structure omitted for clarity: A diatomic molecule AB in the neutral ground state is excited by absorption of electromagnetic radiation at an angular frequency $\omega_{1}$ yielding a population $\rho'$ of excited molecules AB*. These excited molecules AB* are then ionized by electromagnetic radiation at another frequency $\omega_{2}$ forming the molecular ions AB$^{+}$. 
In the case of excitation with linear $z$-polarized radiation, only transitions without change in the projection angular momentum quantum number, i.e., with $M_{N}'=M_{N}''$, are allowed. Therefore, the entire excited state population $\rho'$ is confined to the $M_{N}'=0$ Zeeman state. As a consequence, ionization may only occur from this particular Zeeman state and transitions to AB$^{+}$ from other AB* Zeeman states do not contribute to the ionization process.

In the following, we develop a model for ionization of anisotropically populated levels based on weighting of Zeeman states by their population. 
First, we only  consider spin-rotational fine structure, thereafter we extend our model to cover hyperfine structure as well.

\subsection{Fine structure}

Denoting the population of excited molecules AB* in a certain Zeeman state by $\rho'(J',M'_J)$, the quantity $P_{\rho'}(J',\Jp)$, proportional to the photoionization transition probability between fine-structure levels, is given by
\begin{align}
P_{\rho'}(J',\Jp)  = &
\sum _l \sum _{m_l}  \sum _{m_s} \sum _{M'_J,\MJp}  \rho'(J',M'_J) \nonumber \\
&\times  \Big | \big( \Bra{\np \Lp, \vp, \Np \Lp \Sp \Jp \MJp}  \nonumber \\ 
&\times  \Bra{s,m_s}  \Bra{l,m_l} \big) \boldsymbol{\myhat\mu} \Ket{n' \Lambda', v', N' \Lambda' S' J' \MJ'} \Big |^2.
 \label{eq_anisotrop_fs_transition_probability}
\end{align}

Substituting the matrix element from Eq.~(22) in our previous paper \partOne, we obtain
\begin{widetext}
\begin{align}
P_{\rho'}(J',\Jp) = \: &(2N'+1)(2\Np+1) (2S'+1) (2J'+1)(2\Jp+1)
\sum _l \sum _{m_l}  \sum _{m_s} \sum _{M'_J,\MJp}  \rho'(J',M'_J)
\nonumber \\ 
& \times \left |   \vphantom{\ninej{\Jp}{u}{J}{\Np}{k}{N}{\Sp}{s}{S}}
\sum _{k=l\pm 1} (-1)^{k} \sqrt{2k+1} \tj{l}{ 1}{k}{ -m_l}{ \mu_0}{ -p} \Braket{\np \Lp, \vp | \formsphtMolFixed{k}{\Delta\Lambda} | n' \Lambda', v'} \tj{\Np}{k}{N'}{-\Lp}{\Delta\Lambda}{\Lambda'} 
\right .
\nonumber \\
& \times \left .
\sum_{u=|k-s|}^{k+s}(2u+1) \tj{\Jp}{u}{J'}{-\MJp}{\upr}{\MJ} \tj{u}{k}{s}{\upr}{-p}{m_s}  \ninej{\Jp}{u}{J'}{\Np}{k}{N'}{\Sp}{s}{S'}
\right |^2
 \label{eq_anisotrop_fs_transition_probability_2}
.
\end{align}
\end{widetext}

As the terms in the sum over $M'_{J}$ are weighted by the populations $\rho'(J',M'_J)$, the orthogonality properties of the Wigner 3j-symbols may not be used to eliminate the cross terms in the above expression as is possible for direct ionization (see \partOne). Hence, the vibronic transition matrix elements $\Braket{\np \Lp, \vp | \formsphtMolFixed{k}{\Delta\Lambda} | n \Lambda, v}$ may not be isolated from the other terms in Eq.~\eqref{eq_anisotrop_fs_transition_probability_2} and, since these matrix elements are in general complex quantities, the transition probability may not be calculated unless the magnitude and the (relative) \emph{phases} of these matrix elements are known. In other words, for ionization of an anisotropically populated level, interference effects between different vibronic transition matrix elements become important.\cite{allendorf89a,reid91a}

In order to illustrate this effect, we study the transition probability for a $^{1}\Sigma \rightarrow \hspace{0mm} ^{2}\Sigma$ ionization process for the spin-rotation levels $N'=J'=2 \rightarrow \Np=2, \, \Jp=3/2$ and $N'=J'=2 \rightarrow \Np=2, \, \Jp=5/2$. We assume the population of the neutral $J'=2$ level to be confined to the $M_{J}'=0$ Zeeman state, i.e., $\rho'(J'=2, M_{J}'=0) = 1$ and $\rho'(J'=2, M_{J}' \neq 0) = 0$, as it results from two-photon excitation from the rovibronic ground state with linearly polarized radiation. When writing the vibronic transition matrix elements as complex numbers in polar form, \footnote{The square root in the magnitude of the vibronic transition matrix element has been introduced in order that the relation $C_{k} = \left | \Braket{\np \Lp, \vp | \formsphtMolFixed{k}{\Delta\Lambda} | n' \Lambda', v'} \right | ^2$ still holds.}
\begin{subequations}
\begin{align}
\Braket{\np \Lp=0, \vp | \formsphtMolFixed{k=0}{0} | n' \Lambda'=0, v'} = \sqrt{C_{0}} \exp({i \phi_{0}}),
\\
\Braket{\np \Lp=0, \vp | \formsphtMolFixed{k=2}{0} | n' \Lambda'=0, v'} = \sqrt{C_{2}} \exp({i \phi_{2}}),
\end{align}
\end{subequations}
with $C_{0}, C_{2} \in \mathbb{R}, C_{0}, C_{2} \ge 0$ and $\phi_{0},\phi_{2} \in [0,2\pi)$ and assuming the matrix elements with $k>2$ to vanish,
evaluation of Eq.~\eqref{eq_anisotrop_fs_transition_probability_2} yields
\begin{align}
P_{\rho'}(J'=2,\Jp=3/2) = \: & 0.13 \, C_{0} + 0.07 \, C_{2}
\nonumber \\ 
& - 0.11 \sqrt{C_{0} C_{2}} \cos(\phi_{0} - \phi_{2})
\end{align}
and
\begin{align}
P_{\rho'}(J'=2,\Jp=5/2) = \: & 0.20 \, C_{0} + 0.10 \, C_{2}
\nonumber \\ 
&- 0.16 \sqrt{C_{0} C_{2}} \cos(\phi_{0} - \phi_{2}).
\end{align}

\begin{figure}[tbp]
\begin{center}
\includegraphics[scale=\mathematicaFig]{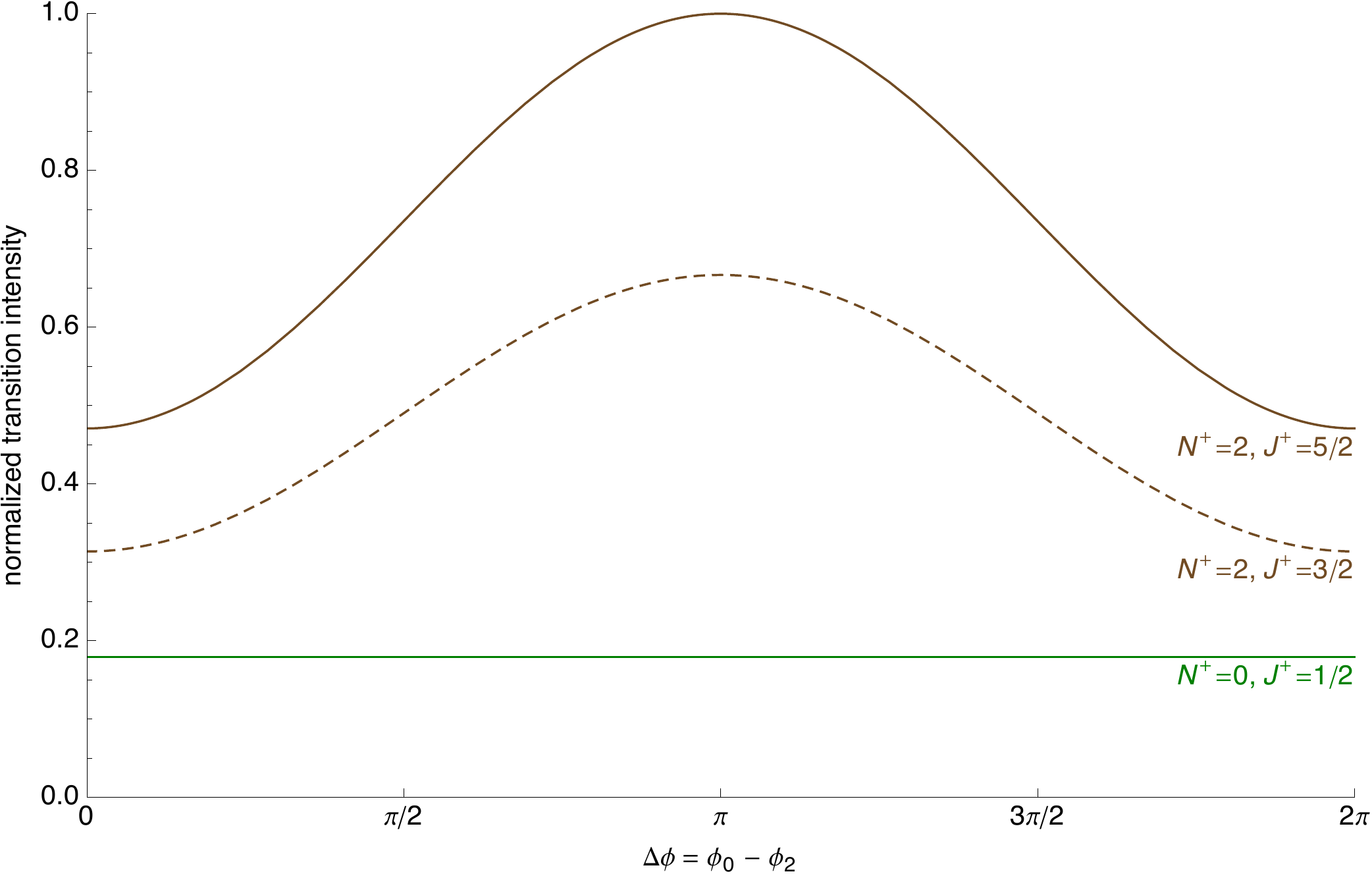}
\caption{
{\bf Interference effects in photoionization of anisotropically populated states.}
When ionizing a neutral molecule from an anisotropically populated level, interference effects between different vibronic transition matrix elements may occur. The intensity of different ionization transitions then not only depends on the magnitude of the vibronic transition matrix elements, but also on their relative phase. Here, this effect is shown for the  $N'=2 \rightarrow \Np=2, \Jp=5/2$ (solid, brown line) and the $N'=2 \rightarrow \Np=2, \Jp=3/2$ (dashed, brown line) transitions, when assuming the entire neutral population being confined to the $N'=2, M_{N}'=0$ Zeeman state. Because of the interference of the vibronic matrix elements with $k=0$ and $k=2$, the transition probability depends on the relative phase $\Delta \phi$ between these matrix elements. For the  $N'=2 \rightarrow \Np=0, \Jp=1/2$ transition (solid, green line), on the contrary, no interference effects are observed, as this transition may only occur due to the $k=2$ vibronic transition matrix element. (For the values shown, the vibronic coefficients have been assumed as $C_{0} = 0.8$, $C_{2} = 0.2$ and all other vibronic coefficients are supposed to vanish.)
}
\label{fig:transition_probability_vs_phase_shift}
\end{center}
\end{figure}

The transition probability for ionization of an anisotropically populated level thus depends not only on the magnitude, but also on the relative phase $\Delta\phi = \phi_{0} - \phi_{2}$ of the vibronic transition matrix elements. This effect is illustrated in Fig.~\ref{fig:transition_probability_vs_phase_shift}. The relative strength of the ionization transitions may therefore in general not be calculated without information about these phases. \footnote{The phases $\phi_{0}$, $\phi_{2}$ of the vibronic transition matrix elements can be related to the scattering phases of the photoelectron. See, e.g., Eq.~(9) and (10) in Ref.~\onlinecite{xie92a}.}

Owing to the non-vanishing cross terms, Eq.~\eqref{eq_anisotrop_fs_transition_probability_2} may in general also not be substantially simplified. Provided the vibronic transition matrix elements are fully specified in terms of both their magnitude and phase (e.g., as a result of an ab-initio calculation), calculation of the quantity $P_{\rho'}(J',\Jp)$ is most conveniently achieved by evaluation of this equation with the help of a computer algebra system.

In many practical relevant cases, however, further simplification of Eq.~\eqref{eq_anisotrop_fs_transition_probability_2} is possible. For photoionization by ejection of the photoelectron from a molecular orbital with predominantly s-type character (such as in H$_2$\cite{merkt92a}, N$_2$\cite{oehrwall99a}, or O$_2$~~\cite{palm98c}), the ionization process is dominated by the vibronic transition matrix elements with the two lowest possible values for $k$, i.e., $k=0$ and $k=2$ for parity conserving transitions,
while matrix elements with higher values of $k$ essentially vanish and may be neglected. Under these conditions, ionizing transitions with a change in the orbital-rotational angular momentum ($\Delta N \neq 0$, i.e., O ($\Delta N =-2$) and S ($\Delta N =+2$) lines),  may only occur due to the $k=2$ vibronic matrix element. Hence, only the $k=2$ term in Eq.~\eqref{eq_anisotrop_fs_transition_probability_2}  is relevant for these transitions. As a consequence, there are no cross terms occurring in that equation and hence also no phase dependencies are observed.
This phase insensitivity is shown in Fig.~\ref{fig:transition_probability_vs_phase_shift} by the example of the $N'=J'=2 \rightarrow \Np=0, \, \Jp=1/2$ transition.

Indeed, the S- and O-lines are the most relevant ones for the state-selective production of molecular cations by the method of threshold REMPI \cite{mackenzie95a,tong10a,tong11a} and we may thus treat these important cases even in absence of information on the phases of the different vibronic matrix elements.

Evaluation of Eq.~\eqref{eq_anisotrop_fs_transition_probability_2} for S and O lines, when taking into account the above-mentioned assumptions and approximations, yields
\begin{widetext}
\begin{align}
P_{\rho'}^\text{(S, O)}(J',\Jp) = \: & 5 (2N'+1)(2\Np+1) (2S'+1) (2J'+1)(2\Jp+1)
\sum _l \sum _{m_l}  \sum _{m_s} \sum _{M'_J,\MJp}  \rho'(J',M'_J)
\nonumber \\ 
& \times
\left | \Braket{\np \Lp=0, \vp | \formsphtMolFixed{k=2}{0} | n' \Lambda'=0, v'} \right | ^2 \tj{l}{ 1}{2}{ -m_l}{ \mu_0}{ -p}^{2} \tj{\Np}{2}{N'}{0}{0}{0}^{2} 
\nonumber \\ 
& \times
 \left |
\sum_{u=3/2}^{5/2}(2u+1) \tj{\Jp}{u}{J'}{-\MJp}{\upr}{\MJ} \tj{u}{2}{1/2}{\upr}{-p}{m_s}  \ninej{\Jp}{u}{J'}{\Np}{2}{N'}{\Sp}{1/2}{S'}
\right |^2
,
 \label{eq_anisotrop_fs_transition_probability_S-O-Lines}
\end{align}
\end{widetext}
where we have also used that $\Lp=\Lambda=0$ for $\Sigma$ states.

\subsection{Hyperfine structure}

The effects discussed so far for ionizing transitions connecting spin-rotational levels are analogously found for hfs-resolved lines. The transition probability for photoionization of excited molecules with populations $ \rho'(F',M'_F)$ for the different Zeeman states is given by:
\begin{widetext}
\begin{align}
P_{\rho'}(J',F',\Jp,\Fp) = \sum _l \sum _{m_l}  \sum _{m_s} \sum _{M'_F,\MFp}  \rho'(F',M'_F)
\Big | 
& \left ( \Bra{\np \Lp, \vp, \Np \Lp \Sp \Jp \Ip \Fp \MFp}  \Bra{s,m_s}  \Bra{l,m_l} \right )
\nonumber \\ & \times
 \boldsymbol{\myhat\mu}  \Ket{n' \Lambda', v', N' \Lambda' S' J' I' F' M'_F}
 \Big |^{2}  .
\end{align}
In a similar way as shown for Eq.~\eqref{eq_anisotrop_fs_transition_probability_2} above, we obtain
\begin{align}
P_{\rho'}(J',F',\Jp,\Fp) = \:  & (2N'+1)(2\Np+1) (2S'+1) (2J'+1) (2\Jp+1) (2F'+1)(2\Fp+1)  \kd{I'}{\Ip}
 \nonumber \\ 
&\sum _l \sum _{m_l}  \sum _{m_s} \sum _{M'_F,\MFp}  \rho'(F',M'_F)
\left | \vphantom{\ninej{\Jp}{u}{J'}{\Np}{k}{N'}{\Sp}{s}{S'}}
\sum _{k=l\pm 1} (-1)^{k} \sqrt{2k+1} 
\right .
\tj{l}{ 1}{k}{ -m_l}{ \mu_0}{ -p} \Braket{\np \Lp, \vp | \formsphtMolFixed{k}{\Delta\Lambda} | n' \Lambda', v'}  
\nonumber \\ & \times
\tj{\Np}{k}{N'}{-\Lp}{\Delta\Lambda}{\Lambda'}
\sum_{u=|k-s|}^{k+s} (-1)^{-u}(2u+1) \tj{u}{k}{s}{\upr}{-p}{m_s}
  \ninej{\Jp}{u}{J'}{\Np}{k}{N'}{\Sp}{s}{S'}
\sj{u}{J'}{\Jp}{I'}{\Fp}{F'}
 \nonumber \\ & \times
\left .
 \tj{\Fp}{u}{F'}{-\MFp}{\upr}{M'_F}
\vphantom{\ninej{\Jp}{u}{J'}{\Np}{k}{N'}{\Sp}{s}{S'}} \right |^2 
 ,
\label{eq_anisotrop_hfs_transition_probability}
\end{align}

upon substituting the results from Eq.~(29), (30) and (22) of our preceding \partOne. As before, evaluation of Eq.~\eqref{eq_anisotrop_hfs_transition_probability} requires in general both the magnitudes and the phases of the vibronic transition matrix elements. Hence, a significant ``paper-and-pencil'' simplification of Eq.~\eqref{eq_anisotrop_hfs_transition_probability} is not possible.

Once more, however, the experimentally interesting case of ionization  out of predominantly s-type orbitals via S and O transitions can be treated even without concrete knowledge of the vibronic transition matrix elements. According to the reasoning above, Eq.~\eqref{eq_anisotrop_hfs_transition_probability} becomes for these transitions:
\begin{align}
P_{\rho'}^\text{(S, O)}(J',F',\Jp,\Fp) 
= \: &
5 (2N'+1)(2\Np+1) (2S'+1) (2J'+1) (2\Jp+1) (2F'+1)(2\Fp+1) \kd{I'}{\Ip}
\nonumber \\ & \times
\left | \Braket{\np \Lp=0, \vp | \formsphtMolFixed{k=2}{0} | n' \Lambda'=0, v'} \right |^2  \tj{\Np}{2}{N'}{0}{0}{0}^2 
 \sum _{l=1,3} \sum _{m_l}  \sum _{m_s} \sum _{M'_F,\MFp}  \rho'(F',M'_F)
\nonumber \\ & \times
\tj{l}{ 1}{2}{ -m_l}{ \mu_0}{ -p} ^2
\left |
 \sum_{u=3/2}^{5/2} (-1)^{-u}(2u+1) \tj{u}{2}{1/2}{\upr}{-p}{m_s}
 \ninej{\Jp}{u}{J'}{\Np}{2}{N'}{\Sp}{1/2}{S'}
\sj{u}{J'}{\Jp}{I'}{\Fp}{F'}
\right .  
 \nonumber \\ & \times
 \left .
 \vphantom{  \ninej{\Jp}{u}{J'}{\Np}{2}{N'}{\Sp}{1/2}{S'} }
 \tj{\Fp}{u}{F'}{-\MFp}{\upr}{M'_F}
\right |^2
.
\label{eq_anisotrop_hfs_transition_probability_SO_Lines}
\end{align}
Further simplification for a singlet neutral excited state ($S'=0$) and for linear polarized radiation ($\mu_0=0$) yields
\begin{align}
P_{\rho'}^{\text{(S, O}, \, S'=0, \text{lin.})} 
(N',F',\Jp,\Fp) 
= 
\: & \frac{5}{2} (2\Np+1)(2N'+1)(2\Jp+1)(2\Fp+1)(2F'+1)\kd{I'}{\Ip}
\tj{\Np}{2}{N'}{0}{0}{0}^2
\nonumber \\  & \times
 \left | \Braket{\np \Lp=0, \vp | \formsphtMolFixed{k=2}{0} | n' \Lambda'=0, v'} \right |^2 
 \sum _{l=1,3} \sum _{m_l}  \sum _{m_s} \sum _{M'_F,\MFp}  \rho'(F',M'_F)
\nonumber \\ & \times
 \tj{l}{ 1}{2}{ -m_l}{ 0 }{ m_l } ^2
\Bigg | \sum_{u=3/2}^{5/2} (2u+1)
\tj{u}{2}{1/2}{-m_l-m_s}{m_l}{m_s}
\sj{\Jp}{u}{N'}{2}{\Np}{1/2}
\nonumber \\ & \times
\sj{u}{N'}{\Jp}{I'}{\Fp}{F'} \tj{\Fp}{u}{F'}{-\MFp}{-m_l-m_s}{M'_F} \Bigg | ^{2} .
\label{eq_anisotrop_hfs_transition_probability_SO_Lines_Singlet}
\end{align}
\end{widetext}


\section{The [2+1'] REMPI process}
\label{sec:REMPI}

Having discussed both, the excitation as well as the ionization step, we may now combine the results from Sec.~\ref{sec:Excitation_step} and \ref{sec:Ionization_step} to a model for the [2+1'] REMPI process.

The populations $\rho'(F', M'_F)$ of the hfs levels in the neutral excited state generated in the REMPI process are proportional to the excitation rates $R_{F'' \rightarrow F', M'_F }$ of the transitions populating these levels multiplied with the population in the relevant levels of the neutral vibronic ground state $\rho''(F'')$:
\begin{equation}
\rho'(F', M'_F)
\propto 
\sum_{F''=|J''-I''|}^{J''+I''}
R_{F'' \rightarrow F', M'_F } \,\rho''(F'')
.
\label{eq_pop_excited_general}
\end{equation}
Here, the sum over all the ground-state hfs levels has been included since hyperfine structure is supposed to be unresolved in the two-photon excitation step. The relative hfs populations of the electronic-ground-state molecules AB are given by a Boltzmann distribution,
\begin{equation}
\rho''(F'') = g_{F''} \exp{(- E_{F''} / k\ri{B} T)},
\label{eq_GS_pops}
\end{equation}
with $E_{F''}$ the energy of the level $F''$ and $g_{F''}$ its degeneracy, $k\ri{B}$ the Boltzmann constant and the $T$ temperature of the thermal ground-state sample. Since the hfs splittings are usually small compared to the thermal energies ($\Delta E\ri{hfs} \ll k\ri{B}T$), the energies $E_{F''}$ are essentially equal for all hfs levels of one rotational level and the exponential factor in Eq.~\eqref{eq_GS_pops} is nearly identical for them. The relative hfs populations within the same rotational level are thus approximately given by the degeneracy of the hfs levels: $\rho''(F'') \propto g_{F''}$. The excitation rate $R_{F'' \rightarrow F', M'_F }$, on the other hand, is proportional to the two-photon line strength worked out in Sec.~\ref{sec:Excitation_step} normalized by the respective ground-state degeneracy $g_{F''}$:
\begin{equation}
R_{F'' \rightarrow F', M'_F }
\propto
1/g_{F''} \times S^\text{(lin)}_\text{S, O}(F'',F',M'_F)
.
\label{eq_pop_excited_specific}
\end{equation}
Combining the results of Eq.~\eqref{eq_pop_excited_general} through \eqref{eq_pop_excited_specific}, we thus obtain the relative hfs populations of the neutral, excited molecules AB* as 
\begin{equation}
\rho'(F', M'_F)
\propto 
\sum_{F''=|J''-I''|}^{J''+I''}
S^\text{(lin)}_\text{S, O}(F'',F',M'_F),
\label{eq_population_propto_summed_LS}
\end{equation}
where $S^\text{(lin)}_\text{S, O}(F'',F',M'_F)$ is the line strength of the two-photon transition in the excitation step given by Eq.~\eqref{eq_two_photon_line_strength_aniso_S_and_O-Lines_lin_pol.} of Sec.~\ref{sec:Excitation_step}.

To obtain the relative populations $\rho^+(\Jp,\Fp)$ of the molecular ions AB$^+$, we substitute the neutral excited state populations $\rho'(F',M'_F)$ in Eq.~\eqref{eq_anisotrop_hfs_transition_probability_SO_Lines} (or Eq.~\eqref{eq_anisotrop_hfs_transition_probability_SO_Lines_Singlet} for a singlet neutral state) by the expression of Eq.~\eqref{eq_population_propto_summed_LS}, i.e,
\begin{equation}
\rho^{+}_{F'}(\Jp,\Fp) 
\propto
P_{\rho'(F', M'_F)'}^\text{(S, O)}(J',F',\Jp,\Fp)
.
\label{eq_ionic_pops}
\end{equation}
Here, the subscript $_{F'}$ has been added indicating that this quantity refers to the ionic population generated via ionization from a particular $F'$ hyperfine level of the neutral excited state. If the hyperfine structure is not resolved in the ionization step, the total ionic population in a particular ionic hyperfine level $\Fp$ is given by the sum over all neutral, excited AB* hfs levels $F'$:
\begin{equation}
\rho^{+}_\mathrm{tot}(\Jp,\Fp)  = \sum_{F'} \rho^{+}_{F'}(\Jp,\Fp).
\label{eq_total_ionic_pops}
\end{equation}
In REMPI experiments, often the two planes of polarization of the excitation and the ionization laser are not parallel, but tilted by some angle $\polarisationAngle$ relative to each other (e.g., as a consequence of the geometry of frequency multiplication stages used for UV generation). If so, the expressions given above for the excitation and the ionization step are referring to two different space-fixed frames of reference.\footnote{For both, the excitation and the ionization step, the polarization of the radiation has been assumed parallel to the $z$-axis of the space-fixed frame. Allowing for an angle between the two polarization vectors thus implies that the calculations for these two steps refer to two different space-fixed frames.}  Such a tilting between the two polarization vectors may be taken into account by multiplication of the populations calculated in the excitation frame (labeled below by projection quantum numbers
$\overline{M}'_F$ as arguments) with a squared Wigner rotation matrix $\WignerD{F'}{\overline{M}'_F}{M'_F}{0}{\polarisationAngle}{0}$ and summing over the projection quantum numbers in the excitation frame $\overline{M}'_F$:\cite{allendorf89a,reid91a}
\begin{align}
\rho'_{\polarisationAngle}(F', M'_F)  = & \sum_{\overline{M}'_F = -F'}^{F'} 
\left [   \WignerD{F'}{\overline{M}'_F}{M'_F}{0}{\polarisationAngle}{0}   \right ]^2 
\rho'(F', \overline{M}'_F)
\label{eq_rotation_of_population}
 \\
 \propto &
\sum_{\overline{M}'_F = -F'}^{F'} 
\left [   \WignerD{F'}{\overline{M}'_F}{M'_F}{0}{\polarisationAngle}{0}   \right ]^2 
\nonumber \\ & \times
\sum_{F''=|J''-I''|}^{J''+I''}
S^\text{(lin. pol.)}_\text{S, O}(F'',F',\overline{M}'_F)
\label{eq_excited_state_populations_rotated}
.
\end{align}
The populations $\rho'_{\polarisationAngle}(F', M'_F) $ are then substituted into Eq.~\eqref{eq_ionic_pops} instead of those obtained from Eq.~\eqref{eq_population_propto_summed_LS} yielding
\begin{equation}
\rho^{+}_{F', \, \alpha}(\Jp,\Fp) 
\propto
P_{\rho'_{\alpha}(F', M'_F)}^\text{(S, O)}(J',F',\Jp,\Fp)
\label{eq_ionic_pops_rotated}
\end{equation}
and
\begin{equation}
\rho^{+}_{\mathrm{tot}, \, \alpha}(\Jp,\Fp)  = \sum_{F'} \rho^{+}_{F', \, \alpha}(\Jp,\Fp).
\label{eq_total_ionic_pops_rotated}
\end{equation}

Although we have concentrated here on the [2+1'] REMPI scheme, our calculations may be adapted for other two-color REMPI schemes as well. In the case of [1+1'] REMPI, the formulae for the excitation step derived in Sec.~\ref{sec:Excitation_step} and then employed in Sec.~\ref{sec:REMPI} must be replaced by the corresponding well-known expressions for one-photon transitions (see, e.g., Refs.~\onlinecite{brown03a,bunker06} and references therein), for [$n$+1'] REMPI (with $n>2$) the excitation step may be treated according to the theory of multiphoton transitions in diatomic molecules discussed in Refs.~\onlinecite{mainos85a,mainos86a,mainos87a}, where also the corresponding formulae for the Hund's case (a) angular momentum coupling scheme as well as for intermediate Hund's case (a)-(b) coupling situations are found. Similarly, the expressions given here for the ionization step may be adapted for other angular momentum coupling cases by means of a suitable basis transformations as, e.g., outlined in Ref.~\onlinecite{brown03a}.


\section{Application: Hyperfine populations of molecular nitrogen ions produced by [2+1'] REMPI}
\label{sec:Application}

\subsection{Non-hfs-resolved photoionization of molecular nitrogen}
 
\subsubsection{S(0)-O(2) REMPI scheme}

\begin{figure}[tbp]
\begin{center}
\includegraphics[scale=0.65]{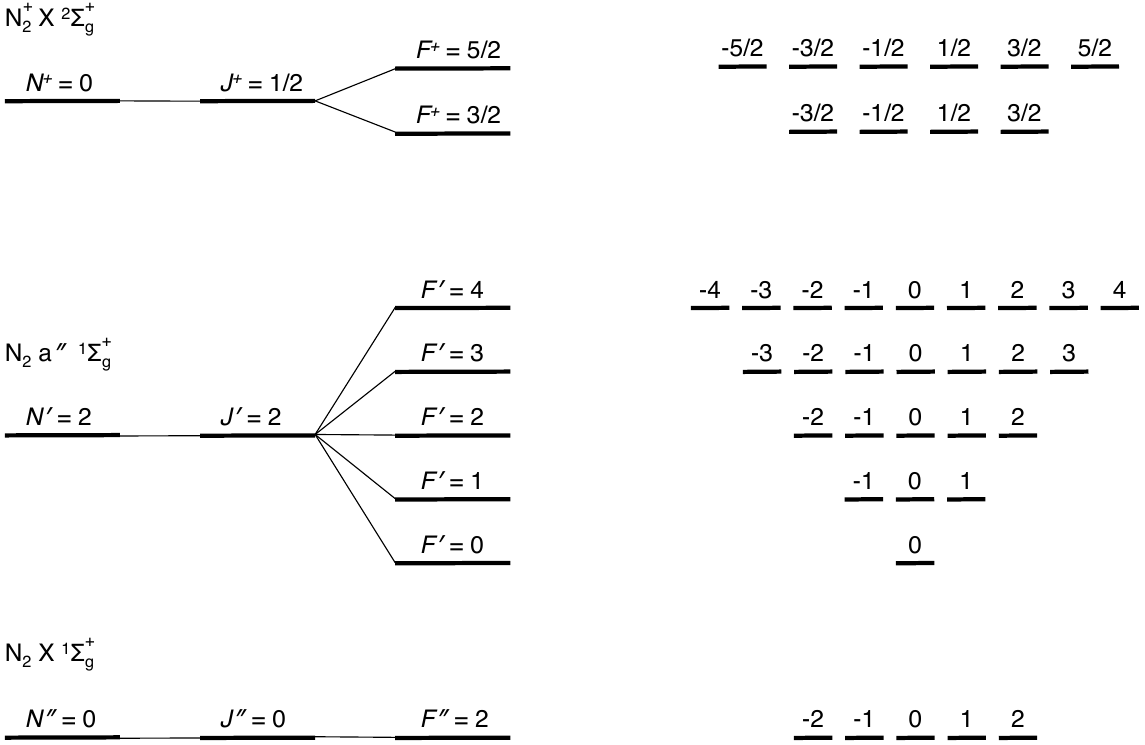}
\caption{
{\bf Level diagram for the S(0)-O(2) REMPI of N$_2$}
showing the relevant fs and hfs levels as well as the Zeeman states for the REMPI sequence $N''=0 \rightarrow N'=2 \rightarrow N^+ =0$ of the $I=2$ nuclear spin manifold of N$_{2}$/\ntp.
(As the exact hyperfine structure of the \ntwoAppState state is unknown, levels of this state are ordered by their degeneracy.)
}
\label{fig:level_diagram_0_2_0}
\end{center}
\end{figure}

As a first application of our model, we study the REMPI scheme previously used for the rotationally state-selective production of \ntp ions via excitation of the N$_2$ \ntwoAppState state.\cite{tong10a,tong11a,tong12a,germann14a} Here, we are interested in the relative hfs populations of \ntp ions produced in the rovibrational ground state by the REMPI sequence $N''=0 \rightarrow N'=2 \rightarrow N^+ =0$ for the $I=2$ nuclear spin manifold of N$_2$/\ntp.
The  energy levels and corresponding Zeeman states involved are shown in Fig.~\ref{fig:level_diagram_0_2_0}.
The hyperfine structure is supposed to be unresolved in both, the excitation and the ionization step. Hence, we calculate the ionic populations using Eq.~\eqref{eq_total_ionic_pops} and Eq.~\eqref{eq_total_ionic_pops_rotated}.

\begin{figure}[htbp]
\begin{center}
\includegraphics[scale=\mathematicaFig]{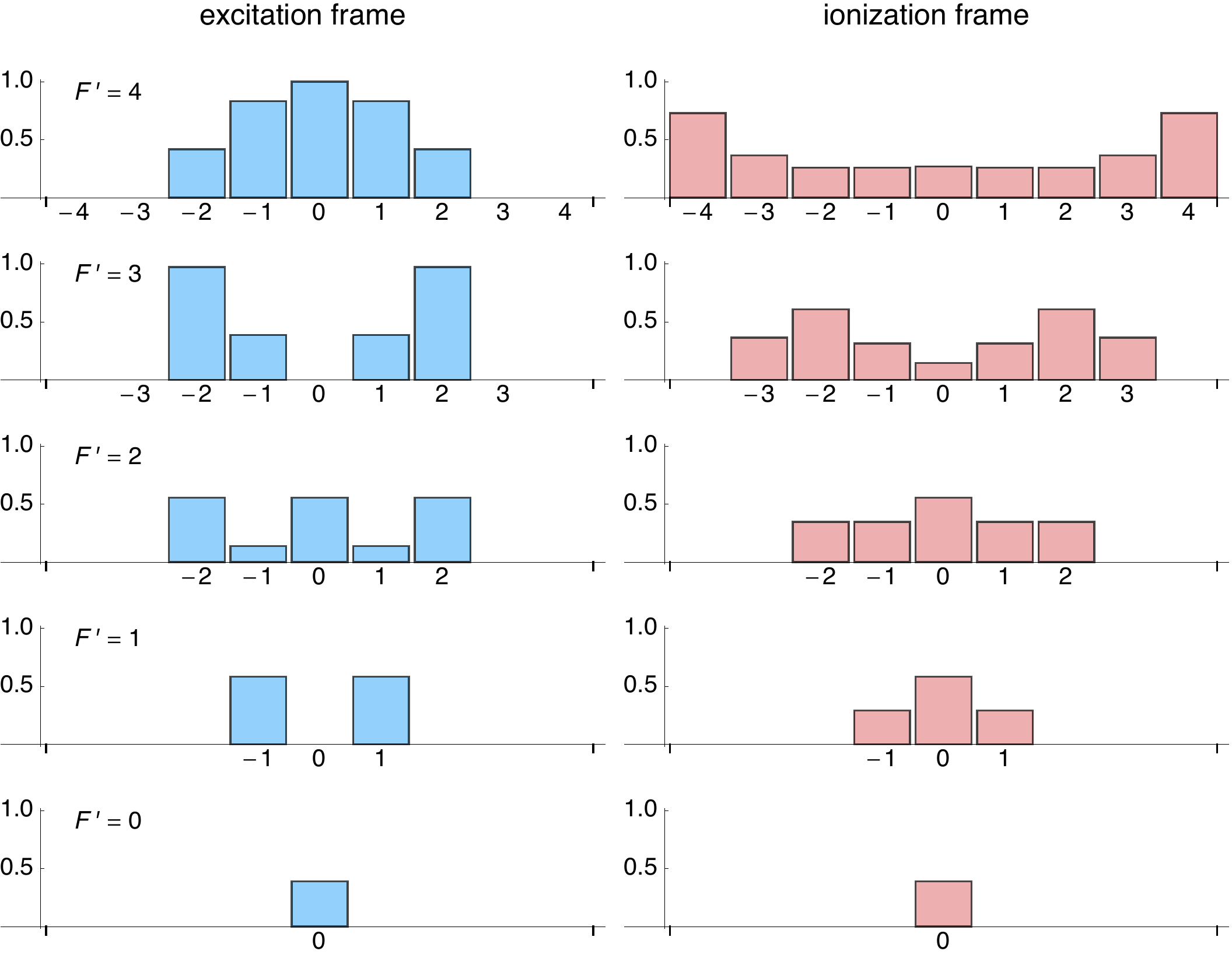}
\caption{
{\bf Populations in the neutral, excited state of the S(0)-O(2) REMPI scheme of N$_2$:} bar charts of the relative populations in the \ntwoAppState state for the REMPI sequence $N''=0 \rightarrow N'=2 \rightarrow N^+ =0$. The  charts in the left column show the relative populations in the Zeeman states of the hfs levels with $F' = 4$ to 0 (top to bottom) with respect to the excitation frame of reference. In the right column, the same populations are shown in the ionization frame of reference, when assuming an angle of $\polarisationAngle = \SI{90}{\degree}$ between the polarization vectors of the excitation and ionization laser beams. The bars are labeled by the projection angular momentum quantum numbers below the horizontal axis.}
\label{fig:excited_populations_0_2_0}
\end{center}
\end{figure}

The relative populations in the neutral, excited N$_2$ \ntwoAppState state are shown in Fig.~\ref{fig:excited_populations_0_2_0}. In the left column of the figure, the populations are shown with reference to the excitation frame, in the right column, the corresponding values after transformation to the ionization frame are given. The effect that only a subset of the Zeeman states may be populated by excitation with linear polarized radiation is clearly visible in the left-hand column of Fig.~\ref{fig:excited_populations_0_2_0}: Zeeman states with $\overline{M}'_{F} > 2$ are not populated due to the selection rule $\overline{M}'_{F} = M''_{F}$ for excitation with linear polarized radiation (polarization vector parallel to quantization axis). 

For the right column of Fig.~\ref{fig:excited_populations_0_2_0}, an angle of $\polarisationAngle = \SI{90}{\degree}$ between the two polarization vectors of excitation and ionization has been assumed. The frame transformation described by Eq.~\eqref{eq_rotation_of_population} leads to a redistribution of the population such that  Zeeman states not populated in the excitation frame, are  populated with respect to the ionization frame of reference.

\begin{figure}[htbp]
\begin{center}
\includegraphics[scale=\mathematicaFig]{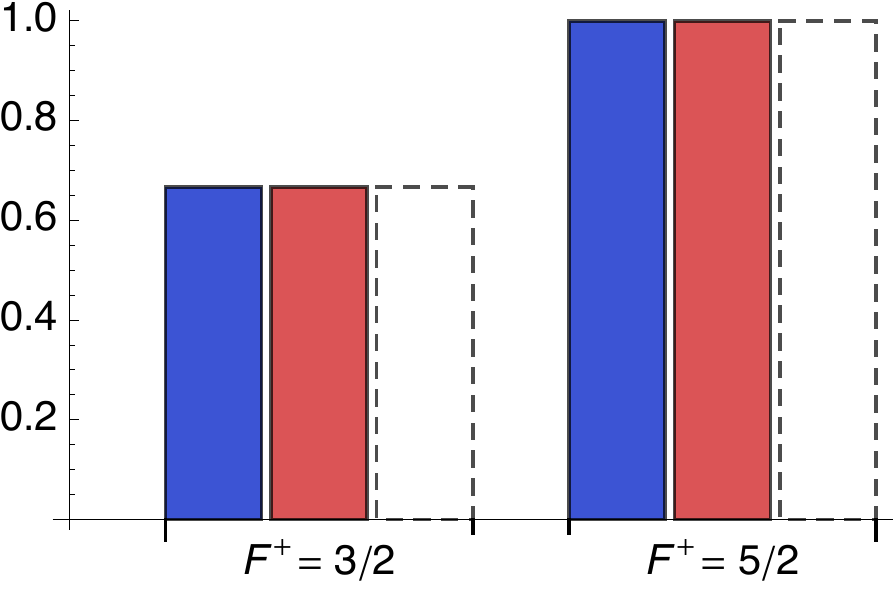}
\caption{
{\bf Populations in the rovibrational ground state of \ntp produced by S(0)-O(2) REMPI:}
relative populations of the two hfs levels of \ntp ions produced by the REMPI sequence $N''=0 \rightarrow N'=2 \rightarrow N^+ =0$.
The blue bars show the populations as obtained for parallel polarization vectors for excitation and ionization ($\polarisationAngle = \SI{0}{\degree}$), for the red ones an angle of $\polarisationAngle = \SI{90}{\degree}$ between the two polarization vectors is assumed.
For reference, the populations expected from the ``pseudo-thermal'' model (see text) are indicated by the white, dash-edged bars. For the example shown, our REMPI model predicts the same relative populations for parallel and perpendicular polarization vectors. Moreover, identical relative populations are also obtained by the ``pseudo-thermal'' model. These coincidences, however, do in general not occur, as illustrated in Fig.~\ref{fig:ionic_populations_2_4_2}. (Values are normalized to yield equal total populations for parallel and perpendicular polarization, as well as for the ``pseudo-thermal'' model, then normalized to unity for the highest value.)
}
\label{fig:ionic_populations_0_2_0}
\end{center}
\end{figure}

The relative ionic populations of the different \ntp hfs levels produced in this REMPI scheme are shown in Fig.~\ref{fig:ionic_populations_0_2_0} as blue and red bars for parallel ($\polarisationAngle = \SI{0}{\degree}$) and perpendicular ($\polarisationAngle = \SI{90}{\degree}$) polarization vectors, respectively. 

For reference, the white, dash-edged bars show the relative populations when assuming them to be proportional to the degeneracy of the levels. We refer to them as the ``pseudo-thermal'' populations, as these are the relative hfs populations of a thermal ensemble in the limit of the thermal energy $k\ri{B}T$ being large compared to the hfs splittings. 

In this particular case, the na\"ive pseudo-thermal model yields the same relative populations as our ionization model does. Moreover, also the relative populations predicted by our model are identical for parallel and perpendicular polarization vectors. As shown below, these coincidences are particular for the S(0)-O(2) ionization sequence and do in general not occur.

\subsubsection{S(2)-O(4) REMPI scheme}

\begin{figure}[htbp]
\begin{center}
\includegraphics[scale=0.65]{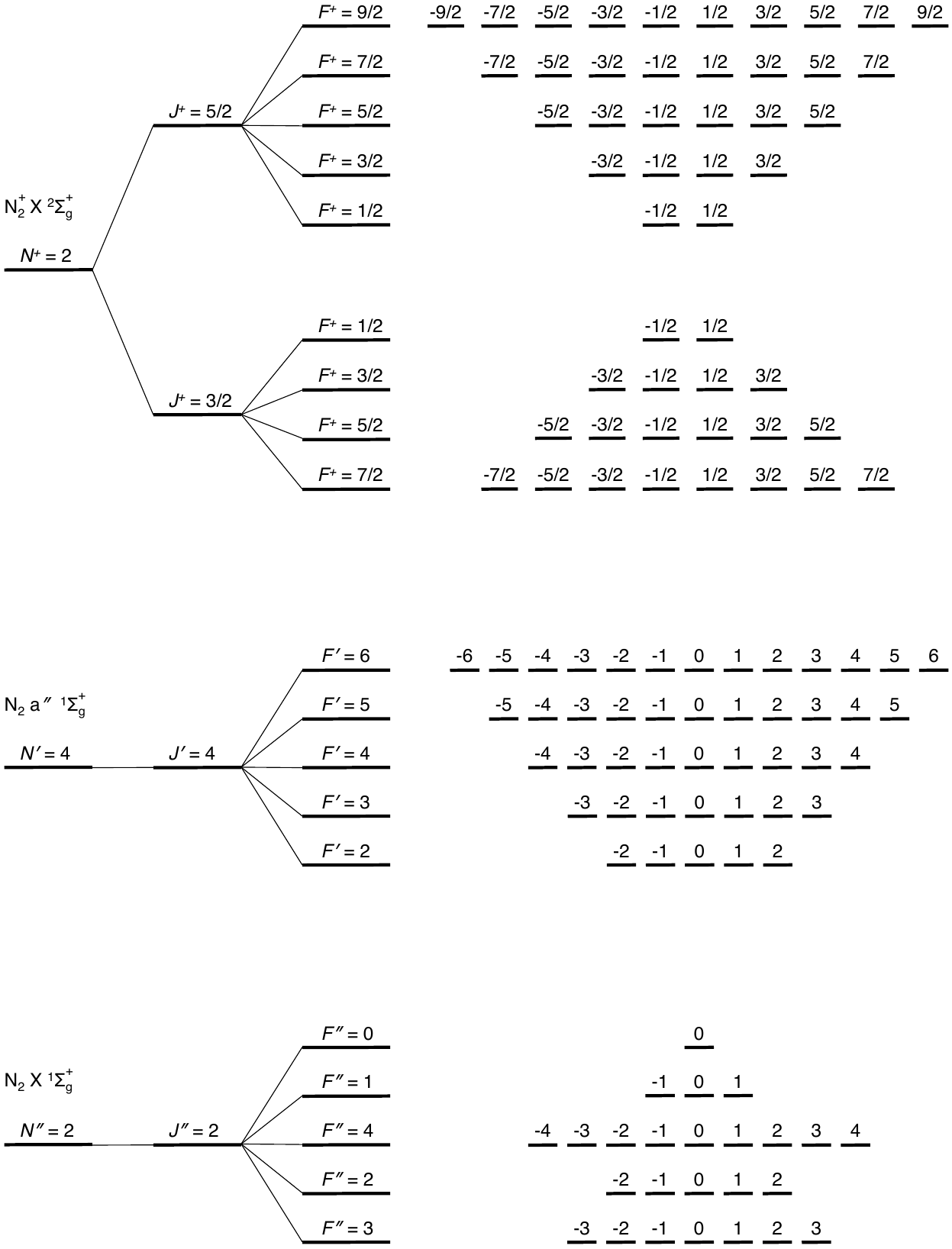}
\caption{
{\bf Level diagram for the S(2)-O(4) REMPI  of N$_2$}
showing the relevant fs and hfs levels as well as the Zeeman states for the REMPI sequence $N''=2 \rightarrow N'=4 \rightarrow N^+ =2$ of the $I=2$ nuclear spin manifold of N$_{2}$/\ntp.
(The energetic order of the hfs levels in the neutral N$_2$ $^1\Sigma_\mathrm{g}^+$ state has been estimated using electric-quadrupole coupling constants extrapolated from spectroscopic data on the neutral N$_2$ A$^3\Sigma^+_\mathrm u$ state and from N$_2$ complexes \cite{freund70a,desantis73a,legon92a}. The hfs levels of the N$_2$ \ntwoAppState are ordered by their degeneracies.)}
\label{fig:level_diagram_2_4_2}
\end{center}
\end{figure}

As a more complex example, we analyze the REMPI of N$_2$ via the \ntwoAppState state through the sequence $N''=2 \rightarrow N'=4 \rightarrow N^+ =2$.
The relevant energy levels and Zeeman states are depicted in Fig.~\ref{fig:level_diagram_2_4_2}. 

\begin{figure}[htbp]
\begin{center}
\includegraphics[scale=\mathematicaFig]{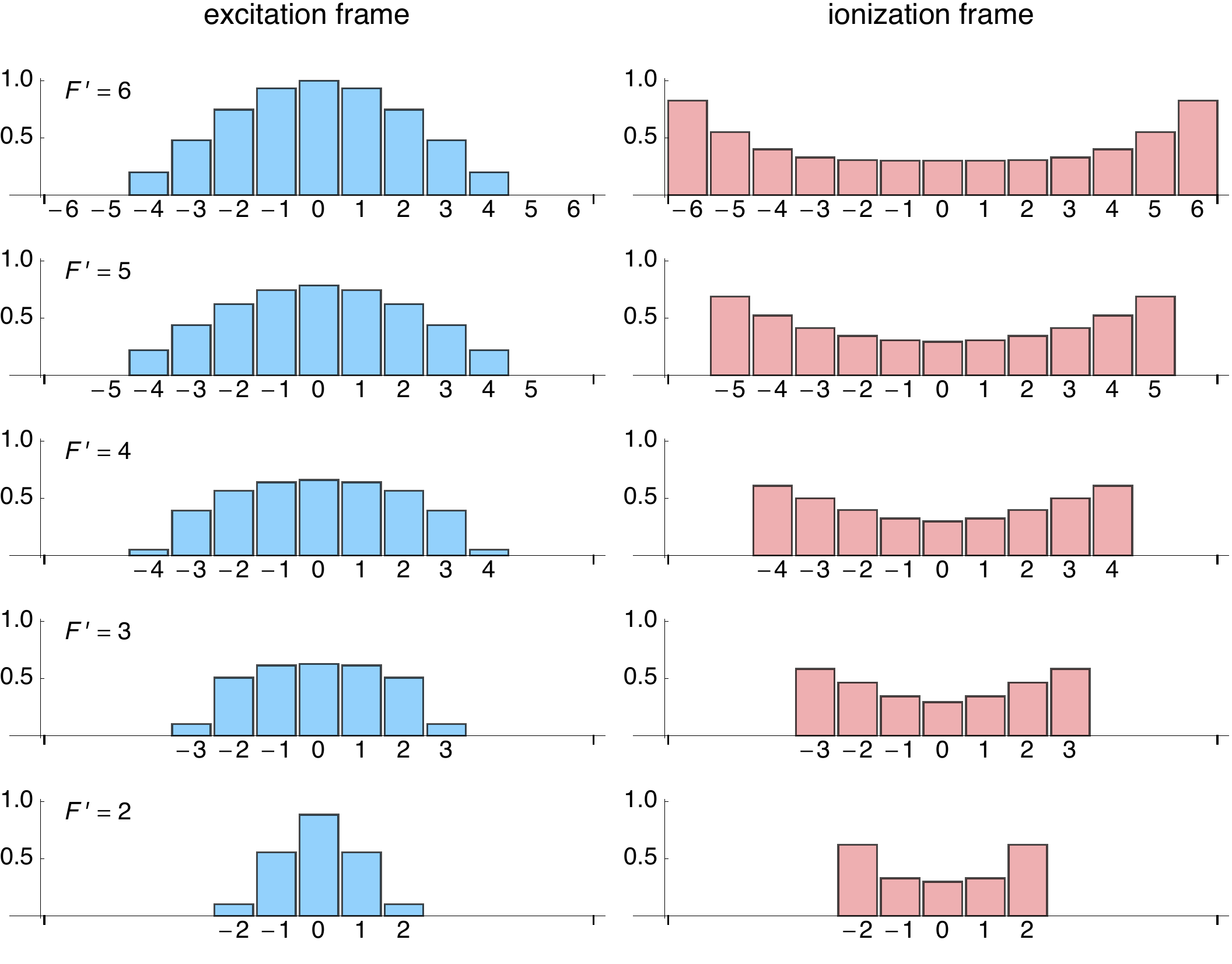}
\caption{ 
{\bf Populations in the neutral, excited state of the S(2)-O(4) REMPI scheme of \ntp:} bar charts of the relative populations in the \ntwoAppState state for the REMPI sequence $N''=2 \rightarrow N'=4 \rightarrow N^+ =2$. The charts in the left column show the relative populations in the Zeeman states of the hfs levels with $F' = 6$ to 2 (top to bottom) with respect to the excitation frame of reference. In the right column, the same populations are shown in the ionization frame of reference, when assuming an angle of $\polarisationAngle = \SI{90}{\degree}$ between the polarization vectors of excitation and ionization. The bars are labeled by the projection angular momentum quantum numbers below the horizontal axis.}
\label{fig:excited_populations_2_4_2}
\end{center}
\end{figure}

\begin{figure}[htbp]
\begin{center}
\includegraphics[scale=\mathematicaFig]{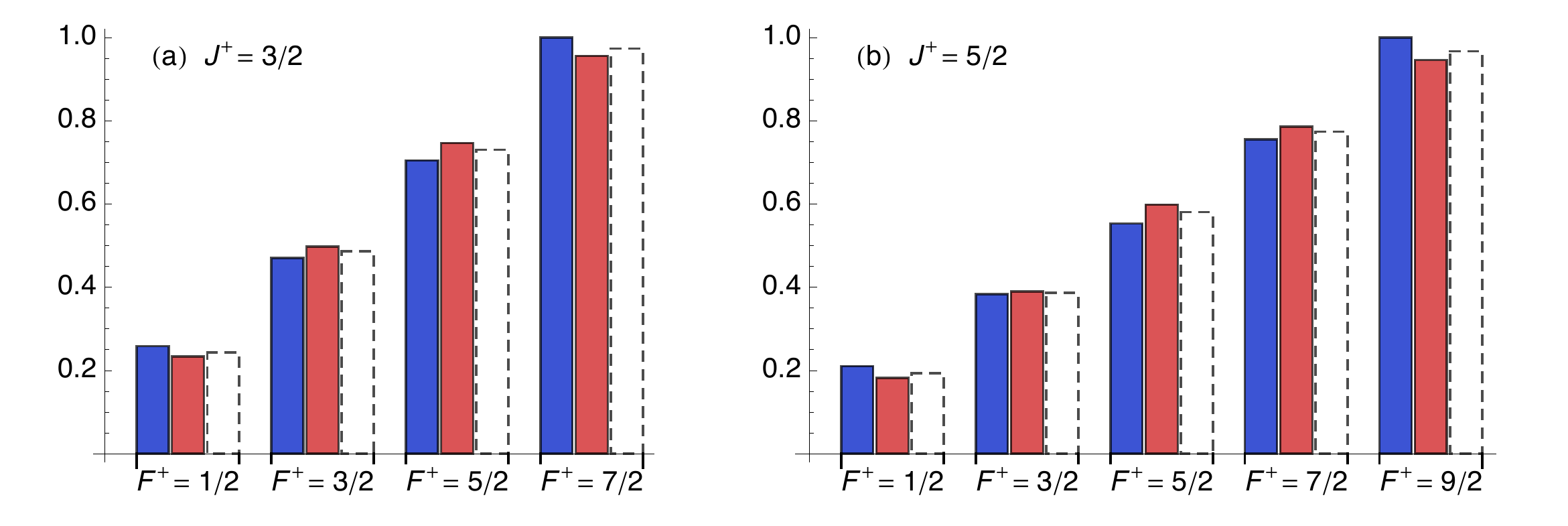}
\caption{
{\bf Ionic populations due to S(2)-O(4) REMPI of \ntp:}
 relative populations of the fs and hfs levels of the $N^+ =2$ rotational state of \ntp produced by the REMPI sequence $N''=2 \rightarrow N'=4 \rightarrow N^+ =2$. {\bf (a)} values for the $\Jp = 3/2$ fine structure level, {\bf (b)} corresponding values for the $\Jp = 5/2$ spin-rotation component. The blue bars show the populations obtained for parallel polarization vectors for excitation and ionization ($\polarisationAngle = \SI{0}{\degree}$), for the red ones an angle of $\polarisationAngle = \SI{90}{\degree}$ between the two polarization vectors has been assumed.
For reference, the populations expected from the pseudo-thermal model (see text) are indicated by the white, dash-edged bars.
}
\label{fig:ionic_populations_2_4_2}
\end{center}
\end{figure}

The populations in the neutral excited \ntwoAppState and the ionic \ntwopXstate state are shown in Fig.~\ref{fig:excited_populations_2_4_2} and Fig.~\ref{fig:ionic_populations_2_4_2}, respectively.
Like in the previous example, the left column of Fig.~\ref{fig:excited_populations_2_4_2} shows the populations in the Zeeman states belonging to the hfs levels of the \ntwoAppState state with respect to the excitation frame of reference, whereas the right column shows them with respect to the ionization frame. As before, an angle  of $\polarisationAngle = \SI{90}{\degree}$ is assumed between the two polarization vectors. Once more, the effect of diminished populations in Zeeman states with high absolute values for the projection angular momentum quantum number due to the selection rules for the excitation step is observed. Also, the redistribution of population in course of the frame transformation described by Eq.~\eqref{eq_rotation_of_population} is seen again.
The relative hfs populations in the $J^+=3/2$ and $J^+=5/2$ spin-rotational levels of the \ntp ion are shown in Fig.~\ref{fig:ionic_populations_2_4_2} (a) and \ref{fig:ionic_populations_2_4_2} (b), respectively.
The populations are shown for both, parallel (blue bars) and perpendicular (red bars) polarization vectors for ionization and excitation. For comparison, also the pseudo-thermal populations are indicated (white, dash-edged bars). 
In contrast to the previous example, the relative hfs populations obtained by our REMPI model now deviate from the pseudo-thermal populations. As these deviations are small, however, they might only have a minor effect on experiments with molecular \ntp ions produced by this and similar REMPI schemes.
We note that the Hund's case \bBetaJ basis states in rotational excited states of \ntp are mixed by off-diagonal terms in the hfs Hamiltonian leading to a mixing of the two fine structure components \cite{berrahmansour91a,germann14a,germann16a}. Since we have found this mixing to only have a minor effect on the hfs populations predicted by our model \cite{germann16a}, we have neglected it here.

\subsection{Hfs-resolved photoionization of molecular nitrogen}

So far, we have analyzed hfs-state populations of \ntp ions generated by hfs-\emph{un}resolved photoionization. This means, ions were assumed to have been produced in a REMPI process, in which the hyperfine structure is not resolved, but the ionic populations were then supposed to be \emph{probed} in a hfs-\emph{resolved} manner, such as by hfs-resolved vibrational spectroscopy of the cation \cite{germann14a}.

Since control over the vibronic and spin-rotational degrees of freedom in the REMPI process has already been achieved \cite{tong10a,tong11a}, extending this state-selectivity to the hfs domain, i.e., producing molecular ions also in a hfs-state-selective manner, is appealing---particularly, in view of emerging non-destructive and coherent  techniques in molecular spectroscopy\cite{mur-petit12a,germann16a, wolf16a}.

Here, we study the implications of our REMPI model for such a hfs-state-selective preparation scheme by analyzing the relative populations of \ntp ions for hfs-resolved ionization transitions. This means, we suppose the same neutral, excited populations as before (Fig.~\ref{fig:excited_populations_0_2_0} and \ref{fig:excited_populations_2_4_2}), but calculate the relative ionic populations for particular $F' \rightarrow \Fp$ transitions using Eq.~\eqref{eq_ionic_pops} and \eqref{eq_ionic_pops_rotated}. We are interested in possible propensities for these hfs-resolved ionization transitions, as these could enable achieving state-selectivity.

\begin{figure}[htbp]
\begin{center}
\includegraphics[scale=\mathematicaFig]{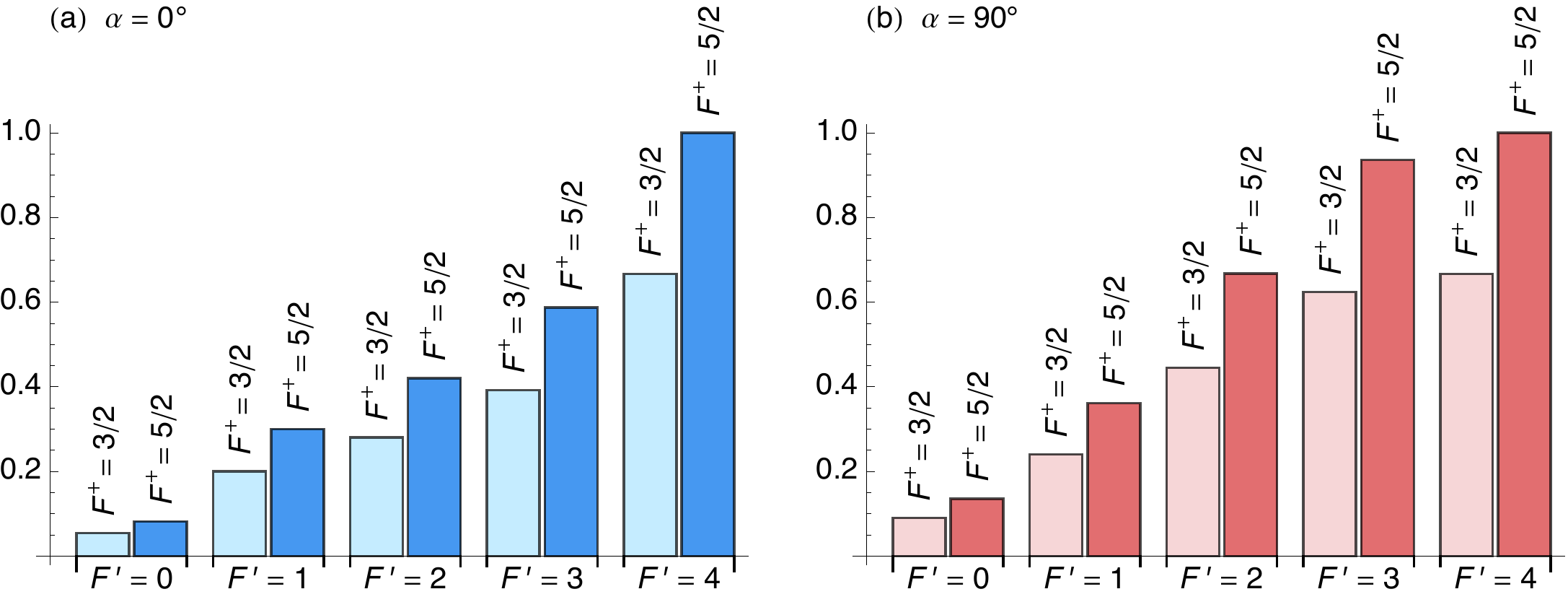}
\caption{
{\bf Populations in the rovibrational ground state of \ntp due to S(0)-O(2) REMPI for specific ionizing hfs transitions:} relative contributions to the two ionic hfs populations $\Fp = 3/2$ and $\Fp = 5/2$ from individual hfs levels of the neutral \ntwoAppState state ($F'= 0, ..., 4$) for parallel (a) and perpendicular (b) polarization vectors. Ionization from all neutral hfs levels populates both ionic hfs states, no clear propensity is observed.
}
\label{fig:ionic_populations_hfs_resolved_0-2-0}
\end{center}
\end{figure}

The results obtained for the S(0)-O(2) REMPI sequence are shown in Fig.~\ref{fig:ionic_populations_hfs_resolved_0-2-0}. As seen from this figure, ionization from all hfs levels of the neutral excited N$_2$ \ntwoAppState state leads to ionic populations in both hfs levels of the rovibronic ground state of \ntp. In other words, no clear propensity is observed.

Hence, hfs-state-selective production of \ntp in the rovibronic ground state would have to be achieved almost entirely by spectroscopic addressing, e.g., by ionizing selectively only above the lowest accessible ionization threshold\cite{tong10a,tong11a}. If this is possible, depends on the bandwidth of the radiation used for ionization and the hfs splitting in the N$_2$ \ntwoAppState state. The latter is unknown at present, since the spectroscopic investigation of this state  \cite{kam91a,lykke91a,vrakking92a,salumbides09a} has not yet achieved hyperfine resolution.

\begin{figure}[htbp]
\begin{center}
\includegraphics[scale=\mathematicaFig]{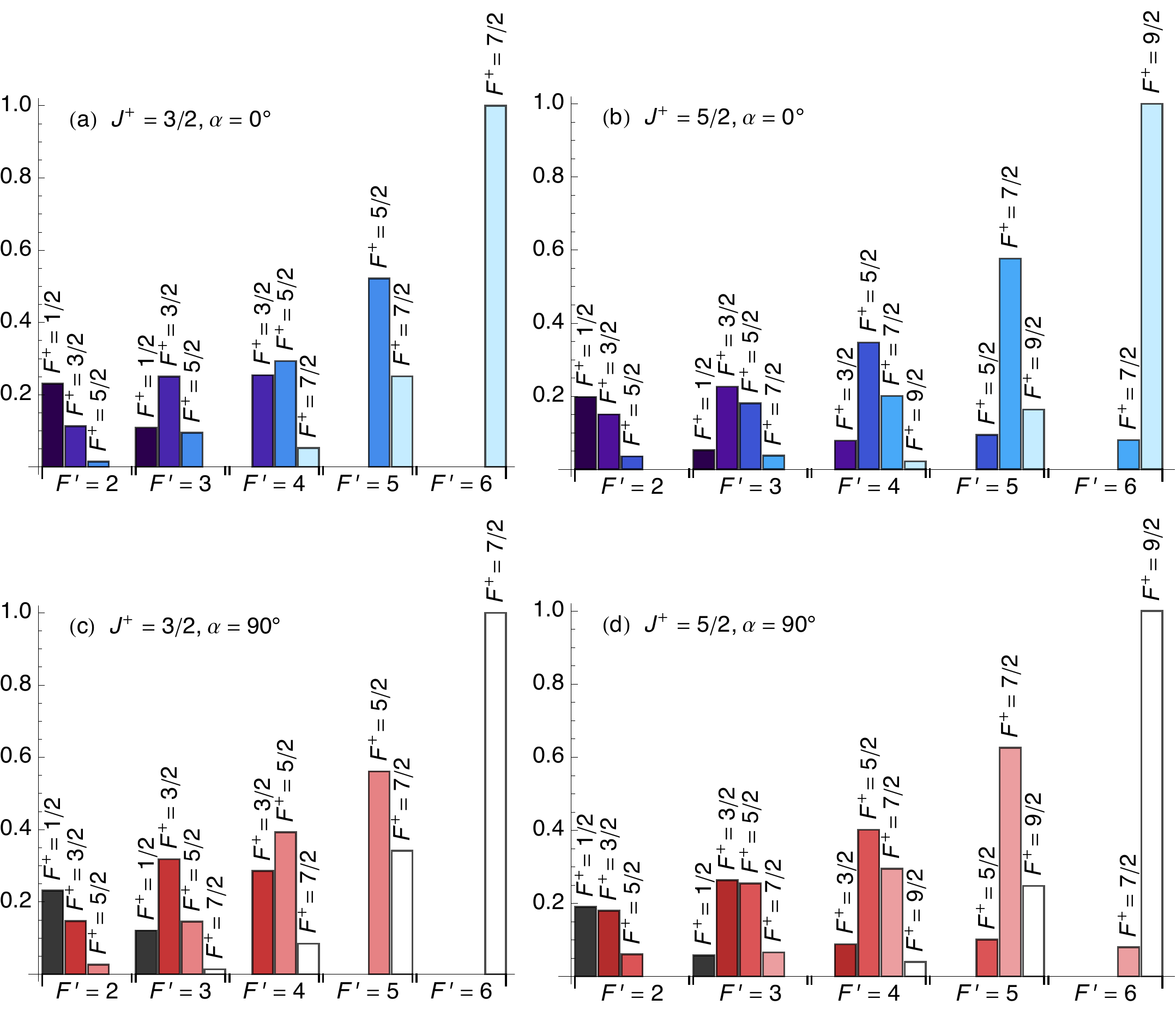}
\caption{
{\bf Ionic populations due to S(2)-O(4) REMPI of \ntp for specific ionization hfs transitions.} 
{\bf Upper row:} contributions to the relative hfs populations of \ntp ions for the $\Jp=3/2$ (panel (a)) and $\Jp=5/2$ (panel (b)) spin-rotation components from individual hfs levels of the neutral \ntwoAppState state assuming parallel polarization vectors of excitation and ionization ($\polarisationAngle = \SI{0}{\degree}$).  {\bf Lower row:} corresponding values for perpendicular polarization vectors ($\polarisationAngle = \SI{90}{\degree}$). For most transitions, a distinct propensity is observed: Ionization from a particular neutral hfs level ($F'=2,...,4$) preferentially results in a population in 
only one, eventually two, ionic hfs levels. (Values $< 10^{-2}$ have been suppressed for clarity.)
}
\label{fig:ionic_populations_hfs_resolved_2_4_2_no_mixing}
\end{center}
\end{figure}

The hfs-resolved results from our model for the S(2)-O(4) REMPI scheme are shown in Fig.~\ref{fig:ionic_populations_hfs_resolved_2_4_2_no_mixing}.\footnote{As before, mixing of Hund's case \bBetaJ basis states has been neglected here for simplicity. Including this mixing has only a minor effect on the results shown (see Ref.~\onlinecite{germann16a}).} For the \ntp ions with $\Np=2$ produced in this scheme, the relative populations exhibit a pattern remarkably different from that seen in the previous example. Ionization via certain hfs levels of the neutral excited \ntwoAppState state populate almost exclusively particular hfs levels in the \ntp ion. In other words, a clear propensity is observed. For the majority of the transitions, this characteristics is summarized by the propensity rule $\Delta J = \Delta F$ (with $\Delta J = \Jp - N'$ and $\Delta F = \Fp - F'$). Deviations from this rule are observed for $\Jp = 3/2$ at low values of $F'$.

As a consequence of this propensity, hfs-state-selectivity can be achieved even without full spectroscopic resolution of individual hfs transitions in the ionization step.

\section{Summary and Conclusions}
\label{sec:Conclusions}

In this paper, we have presented a model for the calculation of the relative populations of fine and hyperfine levels of molecular cations produced by resonance-enhanced multiphoton ionization (REMPI). Our model is based on understanding the REMPI process as two separate steps, a bound-bound neutral-ground-to-neutral-excited-state transition followed by the ionizing transition generating the cation. 

Compared to the model for fine- and hyperfine structure effects in one-photon ionization presented in the preceding article \partOne, description of the REMPI process requires considering two additional effects: hyperfine effects in the neutral bound-bound multiphoton transition and ionization of an anisotropic sample of neutral molecules.

Anisotropy complicates the calculation of transition probabilities and---in general---leads to interference effects between different vibronic transition matrix elements, resulting in a dependence of the observed ionization intensities not only on the magnitudes, but also on the phases of these matrix elements. Prediction of transition intensities hence is in general only possible with vibronic transition matrix elements fully characterized by both their magnitude and their phase. However, in the practically particularly relevant cases of S and O ionization transitions with the photoelectron ejected from a molecular orbital with predominantly s-type character, calculation of the relative populations of fine and hyperfine levels in the cation is possible without such detailed information.

We have shown the implications of our model using the REMPI of molecular nitrogen via the  \ntwoAppState excited state as a representative example. Our results may be used to calculate relative fs and hfs populations in molecular cations produced by REMPI for subsequent spectroscopy or dynamics experiments and may thus assist the interpretation of results obtained from such experiments. Moreover, they may serve as a theoretical background to develop future fine- and hyperfine-state-selective production schemes for molecular cations.


\begin{acknowledgments}
This work has been supported by the Swiss National Science Foundation as part of the National Centre of Competence in Research, Quantum Science \& Technology (NCCR-QSIT), the European Commission under the Seventh Framework Programme FP7 GA 607491 COMIQ and the University of Basel
\end{acknowledgments}

\bibliography{Main-Jan2016_modified,HF_PI_dynamics_refs_reduced_arXiv,additional_Refs}

\end{document}